\documentclass[preprint]{emulateapj}
\usepackage{xspace}
\usepackage{natbib}
\usepackage{verbatim}
\usepackage{graphicx,epstopdf}
\usepackage{longtable}
\usepackage{fancyref}
\usepackage{hyperref}
\usepackage{multirow}
\usepackage{amsmath}
\newcommand{\kep}{{\it Kepler}}
\newcommand{\ktwo}{{\it K2}\xspace}

\newcommand{\vsini}{$v \sin i$}

\newcommand{\nd}{---}

\newcommand{\npc}{\ensuremath{34}\xspace}
\newcommand{\neb}{\ensuremath{184}\xspace} 
\newcommand{\nvar}{\ensuremath{222}\xspace} 
\newcommand{\ntce}{\ensuremath{1274}\xspace}
\newcommand{\nstars}{\ensuremath{\sim}34,000\xspace} 

\providecommand{\adsurl}[1]{\href{#1}{ADS}}
\providecommand{\eprint}[1]{\href{http://arxiv.org/abs/#1}{#1}}
\def\aap{{A\&A}}		
\def\apj{{ApJ}}			
\def\apjs{{ApJS}}		
\def\pasp{{PASP}}		

\def\mnras{{MNRAS}}
\def\nat{{Nature}}

\newcommand{\teff}{$T_\mathrm{eff}$\xspace}
\newcommand{\logg}{$\log g$\xspace}
\newcommand{\kms}{\ensuremath{\rm km\,s^{-1}}}
\newcommand{\ms}{\ensuremath{\rm m\,s^{-1}}}
 
\newcommand{\TERRA}{\texttt{TERRA}\xspace}
\newcommand\nsf{$\dagger$}
\newcommand\nsfp{$\star\star$}
\newcommand\hub{$\star$}
\newcommand\texaco{$\ddagger$}

\slugcomment{}

\shorttitle{K2 C17 Planet Candidates}
\shortauthors{Crossfield et al.}

\begin{document}


\title{A TESS Dress Rehearsal: Planetary Candidates and Variables from \ktwo Campaign 17}


\author{Ian J.\ M.\ Crossfield\altaffilmark{1}, 
  Natalia Guerrero\altaffilmark{1},
  Trevor David\altaffilmark{2},
  Samuel N.\ Quinn\altaffilmark{3},
  Adina D.\ Feinstein\altaffilmark{4},
  Chelsea Huang\altaffilmark{1},
  Liang Yu\altaffilmark{1},
  Karen A.\ Collins\altaffilmark{3},
  Benjamin J.\ Fulton\altaffilmark{5},
Bj{\"o}rn Benneke\altaffilmark{6},
Merrin Peterson\altaffilmark{6},
Allyson Bieryla\altaffilmark{3},
Joshua E.\ Schlieder\altaffilmark{7},
Molly R.\ Kosiarek\altaffilmark{13,\nsf},
Makennah Bristow\altaffilmark{8},
Elisabeth Newton\altaffilmark{1,\nsfp},
Megan Bedell\altaffilmark{9},
David W.\ Latham\altaffilmark{3},
Jessie L.\ Christiansen\altaffilmark{5},
Gilbert A. Esquerdo\altaffilmark{3},
Perry Berlind\altaffilmark{3},
Michael L. Calkins\altaffilmark{3},
Avi Shporer\altaffilmark{1},
Jennifer Burt\altaffilmark{1},
Sarah Ballard\altaffilmark{1},
Joseph E.\ Rodriguez\altaffilmark{3},
Nicholas Mehrle\altaffilmark{1},
Courtney D.\ Dressing\altaffilmark{10},
Sara Seager\altaffilmark{1,11},
Jason Dittmann\altaffilmark{1},
David Berardo\altaffilmark{1},
Lizhou Sha\altaffilmark{1},
Zahra Essack\altaffilmark{11},
Zhuchang Zhan\altaffilmark{11},
Martin Owens\altaffilmark{1},
Isabel Kain\altaffilmark{1},
John H.\ Livingston\altaffilmark{12},
Erik A.\ Petigura\altaffilmark{13,\hub},
Erica J.\ Gonzales\altaffilmark{14,\nsf},
Howard Isaacson\altaffilmark{10},
Andrew W.\ Howard\altaffilmark{13}
}

\altaffiltext{1}{Department of Physics, and Kavli Institute for Astrophysics and Space Research, Massachusetts Institute of Technology, Cambridge, MA 02139, USA}
\altaffiltext{2}{Jet Propulsion Laboratory, California Institute of Technology, 4800 Oak Grove Drive, Pasadena, CA 91109, USA}
\altaffiltext{3}{Center for Astrophysics, 60 Garden Street,  Cambridge, MA 02138, USA}
\altaffiltext{4}{Department of Physics and Astronomy, Tufts University, Medford, MA 02155, USA}
\altaffiltext{5}{Caltech/IPAC-NASA Exoplanet Science Institute, 770 S. Wilson Ave, Pasadena, CA 91106, USA}
\altaffiltext{6}{Departement de Physique, Universite de Montreal, Montreal, H3T 1J4, Canada}
\altaffiltext{7}{NASA Goddard Space Flight Center, 8800 Greenbelt Road, Greenbelt, MD 20771, USA}
\altaffiltext{8}{Department of Physics, University of North Carolina at Asheville, Asheville, NC 28804, USA}
\altaffiltext{9}{Center for Computational Astrophysics, Flatiron Institute, 162 5th Ave., New York, NY 10010, USA}
\altaffiltext{10}{Astronomy Department, University of California, Berkeley, CA, USA}
\altaffiltext{11}{Department of Earth, Atmospheric and Planetary Sciences, Massachusetts Institute of Technology, Cambridge, MA 02139, USA}
\altaffiltext{12}{Department of Astronomy, Graduate School of Science, The University of Tokyo, Hongo 7-3-1, Bunkyo-ku, Tokyo, 113-0033, Japan}
\altaffiltext{13}{Cahill Center for Astrophysics, California Institute of Technology, Pasadena, CA 91125, USA}
\altaffiltext{14}{Department of Astronomy and Astrophysics, University of California, Santa Cruz, CA 95064, USA}
\altaffiltext{\nsf}{NSF Graduate Research Fellow}
\altaffiltext{\nsfp}{NSF Postdoctoral Fellow}
\altaffiltext{\hub}{NASA Hubble Fellow}
\altaffiltext{\texaco}{Texaco Fellow}

\begin{abstract}

We produce light curves for all \nstars\ targets observed with K2 in
Campaign 17 (C17), identifying \npc\ planet candidates,
\neb\ eclipsing binaries, and \nvar\ other periodic variables.  The
forward-facing direction of the C17 field means follow-up can begin immediately now
that the campaign has concluded and interesting targets have been
identified.  The C17 field has a large overlap with C6, so this latest
campaign also offers a rare opportunity to study a large number of
targets already observed in a previous K2 campaign.  The timing of the
C17 data release, shortly before science operations begin with the
Transiting Exoplanet Survey Satellite (TESS), also lets us exercise
some of the tools and methods developed for identification and
dissemination of planet candidates from TESS. We find excellent
agreement between these results and those identified using only
K2-based tools.  Among our planet candidates are several planet
candidates with sizes $<$4$R_\oplus$ and orbiting stars with
$Kp\lesssim10$ (indicating good RV targets of the sort TESS hopes to
find) and a Jupiter-sized single-transit event around a star already
hosting a 6~d planet candidate.




\end{abstract}

\keywords{methods: data analysis, planets and satellites: detection,  techniques: photometric}

\section{Introduction}


Launched in 2009, the success of \textit{Kepler} and its extended
mission, K2, is unprecedented.  In addition to their considerable
contributions to other areas of astrophysics, these missions have led
to planets candidates and confirmed planets in the thousands ({\em
  Kepler}) and hundreds ({\em K2}). Unlike the original
\textit{Kepler} mission, K2 observes along the ecliptic plane,
providing 30-minute-cadence light curves for several thousand targets
in each roughly 80-day campaign \citep[]{howell:2014}.

The surge of data provided by the mission at the end of each campaign
is processed and vetted for potential planet candidates. Due to
spacecraft systematics and various sources of astrophysical variability, systems
showing interesting signals are vetted by-eye before proceeding with
additional confirmation follow-up with ground-based telescopes.


The recently launched Transiting Exoplanet Survey Satellite (TESS)
will observe $\sim$ 90\% of the sky, approximately 400 times what
\textit{Kepler} observed and 26 times what K2 has observed so
far. While experience shows that the vetting of potential planet
candidates from K2 campaigns can be completed by a single person or a
small team, the number of TESS candidates to be sifted may be far
larger. Partly for that reason, TESS employs a larger and
better-funded team that has been preparing a set of advanced
diagnostics and tools.  Because TESS observes in the anti-sun
direction while orbiting the Earth \citep{ricker:2014}, if TESS
candidates can be quickly identified after each sector, they can be
immediately sent to ground-based observers to confirm the planets and
study them in more detail.


The recent delivery of data from K2 Campaigns 16 and 17 (C16 and C17)
have provided us with the chance to exercise some of the tools and
techniques being developed for rapid planet candidate identification
and dissemination from TESS and compare results to previous techniques
used for K2.  We conducted a rapid analysis of data from C16 using
tools and methods developed strictly for K2 \citep{yu:2018b}. With
C17, we include a more TESS-like analysis using several of the tools
and team members that will soon examine real TESS data.

C16 and C17 are also ``TESS-like'' in at least two other ways. First,
these are both ``forward-facing'' campaigns in which the
Earth-trailing K2 observed roughly anti-sun from the Earth; as with
TESS sectors (see above),  K2's forward-facing fields can be
immediately observed from the ground if identified with sufficient
rapidity. Second, both of these fields partial overlap with previous
K2 campaigns: C16 with C5 (observed April--July 2015) and C17 with C6
(July--September 2015).  The rare overlap between C17 and C6 offers an
opportunity to study for again a large number of targets previously
observed by K2.  Campaign~18, currently being observed, will also
partly overlap C5 and C16.  Similarly, repeated observations of the
same targets will occur regularly when TESS begins near-continuous,
year-long observations of the ecliptic poles.

Here, we present the techniques and results of our rapid
identification of planet candidates and other astrophysical variables
observed in C17. Sec.~\ref{sec:methods} details the identification
process of planet candidates using methods and tools developed for
both K2 and for TESS. Stellar and planet candidate parameters are
discussed in Sec.~\ref{sec:params}. Sec.~\ref{sec:results}
discusses the results from the two independent vetting techniques
described in Sec.~\ref{sec:methods}. Similarities and discrepancies
between planet candidates identified in C17 and C6 are discussed in
Sec.~\ref{sec:comparison}. We remark on several
individually-interesting systems in Sec.~\ref{sec:indiv}, and
finally conclude in Sec.~\ref{sec:conclusions}.

\section{Identifying Planet Candidates}
\label{sec:methods}
K2 observed C17 from March~1 until May~8, 2018. At 68 days, the
campaign is slightly shorter than most previous K2 campaigns. We
followed exactly the methods of \cite{yu:2018b} to compute photometry
and identify transit-like Threshold Crossing Events (TCEs).As soon as
the raw cadence files were transferred from the spacecraft and
uploaded to MAST, we downloaded these data and began our analysis.
We converted raw K2 cadence data to target pixel files with
\texttt{kadenza}\footnote{\url{https://github.com/KeplerGO/kadenza}}
\citep{barentsen:2018}, converted pixel files to time-series
photometry with
\texttt{k2phot}\footnote{\url{https://github.com/petigura/k2phot/}},
and identified TCEs in light curves using
\TERRA\footnote{\url{https://github.com/petigura/terra}}
\citep{petigura:2015phd,petigura:2018}. We have uploaded light curves for all
 C17 sources outside the Solar system in machine-readable format on the ExoFOP-K2
website \footnote{\url{https://exofop.ipac.caltech.edu/k2/}}.

We identified \ntce\ TCEs with multi-event statistic (effectively a
measure of signal-to-noise) $\ge10$, and pursued two parallel paths to
winnow down these \ntce\ TCEs to a list of reliable planet
candidates. In one, we used a set of new tools being developed for
efficient and robust vetting of candidates expected to be delivered
soon by TESS; we hereafter refer to this as TESS-like candidate
vetting. We also employed a so-called K2-like vetting approach by
using a set of K2-specific tools and practices that have been refined
through the past four years of K2 operations
\citep{crossfield:2015a,crossfield:2016,crossfield:2017,schlieder:2016,obermeier:2016,sinukoff:2016,petigura:2018,ciardi:2018,david:2018,yu:2018b}. We
outline both approaches below, and later compare the results of each
in \ref{sec:poe}.

\subsection{TESS-Like Vetting}
In this effort we use the \TERRA data products with the TESS Exoplanet
Vetter (TEV), which is the web interface tool developed as part of the
TESS Science Office data pipeline. TEV will be used to identify TESS
Objects of Interest (TOIs) in the TCEs found in the TESS pipeline of
record run by the Science Payload Operations Center (SPOC) at
NASA/Ames and the internal Quick-Look Pipeline (QLP; Huang et al., in
prep.) run at MIT. TEV was developed at MIT by the TESS Science Office
staff, and will be described in more detail by Guerrero et al.\ (in
prep.)

TEV imports a data delivery into a database and displays various
vetting plots and data for the candidate TCEs for the
first round of vetting by individuals. The data reduction pipeline that generated the
analysis products --- in this case \TERRA, but SPOC or QLP for
TESS science operations --- provides an analysis summary page for each
candidate TCE and a more comprehensive multi-page analysis report. The
pipeline also provides a spreadsheet with the EPIC or TIC ID, and
basic stellar and transit parameters.

During the individual vetting phase, human vetters inspect the light
curve and other metrics in the analysis summary page (and extended
report if necessary) to determine whether the candidate is a planet
candidate (PC), eclipsing binary (EB), stellar variability (V), other
astrophysical source of variability (O), instrument or systematic noise
(IS), or undecided (U). For multi-planet systems, the candidates can
be compared consecutively. Each individual vetter assigns a
disposition to the candidate and has the option to make additional
comments about the candidate. To complete the individual vetting
stage, a candidate must get at least three unanimous individual
dispositions or up to five total dispositions.  The K2 C17 delivery
had \ntce\ TCEs. A group of nineteen vetters completed the initial
vetting stage in less than twenty-four hours after the delivery was
imported into TEV.

TCEs classified unanimously as EB, V, or IS are automatically assigned
that value as their final disposition. Targets classified unanimously
as PC or with differing dispositions between vetters are flagged for
group vetting, the second stage of the vetting process.  Once the
initial individual vetting concludes, group vetting begins by
resolving conflicts for systems classified with at least one planet
candidate or undecided disposition.  Following this, the group
inspects TCEs dispositioned unanimously as planet
candidates. Conflicts between EB, V, and IS are resolved last. In this
C17 exercise, the group applied and practices the conventions for
assigning candidate dispositions that will be carried over to nominal
TESS operations, including how to disposition and annotate contact
binaries, candidates in a multi-transit system triggered by an
eclipsing binary's secondary eclipses, and candidates with radii $> 30
R_\oplus$.

The group vetting process took about three hours to disposition 180
TCEs. This duration is not fixed, and is likely to evolve as TESS
vetters are trained.  Systems identified in the exercise as known
planets or eclipsing binaries were still dispositioned as PC, but in
nominal TESS operations, TEV will filter candidates using catalogs of
known planets, eclipsing binaries, and variable stars. Several of the
candidates identified as strong candidates for observation were known
targets in K2's Campaign 6, which demonstrates that TEV users have the
materials and expertise necessary to reliably identify planet
candidates.

At the conclusion of group vetting, a TEV administrator closed the K2
C17 delivery to additional changes and TEV generated the final
disposition list for download by TEV users. As in nominal TESS
operations, the final C17 list was disseminated to the TESS Follow-Up
Observing Program (TFOP\footnote{\url{https://tess.mit.edu/followup/}}).

Although we have endeavored to implement the full TESS vetting
process, our K2 C17 vetting diagnostic products did not provide the
full diagnostic capabilities that will be available from the SPOC and
QLP pipelines for TESS vetting. First, no centroid shift information
was available to aid in identifying nearby eclipsing binaries from the
K2 data alone, on account of K2's extremely high pointing
jitter. Second, the K2 vetting diagnostics provided access to a light
curve from only one photometric aperture per target. TESS pipelines
will provide light curves from several aperture sizes to help to
identify blended EB false positives. Third, the TESS analysis will
implement ephemeris matching between the 2-minute-cadence postage
stamps (a restricted set of targets) and the 30-minute-cadence full
frame images (FFIs)  to provide an additional means of
identifying TESS aperture contamination by near or distant variable
sources; we did not employ ephemeris matching in our C17
vetting. Finally, an extensive catalog of known variables and transit
false positives is under development. TESS TCEs will be automatically
crossed-referenced to data in the catalog before the human vetting
process begins, but since this catalog is not yet complete we did not
cross-reference our C17 candidates against it.

\subsection{K2-Like Vetting}
           
Our K2-like vetting procedure closely followed previous efforts by our
group \citep[e.g.,][]{yu:2018b}. Six participants inspected a subset of
TCEs that were assigned in order of TCE number (the EPIC ID appended
by the candidate number). This pseudo-random scheme ensured that a
given vetter inspected a sample of signals that covered a range of
S/N. Each TCE was inspected by at least one person, and by the end of
the vetting procedure 986 TCEs were inspected by 2 or more people
(with 288 inspected by only one person). This resulted in 2548
individual dispositions for the \ntce\ TCEs, across 87 unique potential 
candidates.

Of these 87 signals, 45 were consistently identified as planet
candidates by at least 2 people and 50 were identified as a candidate
by at least one person without contest. While this vetting procedure
was necessarily subjective, the common characteristics we looked for
in the \TERRA diagnostic plots in order to assign the disposition of a
candidate were: consistent depth, no obvious odd/even variations in
depth or transit time which might suggest an EB, lack of an obvious
secondary eclipse, and lack of significant phase-coherent
out-of-transit variability. We did not penalize signals for being
V-shaped alone. However, if a TCE was deep, V-shaped, and long in
duration yet still lacked an obvious secondary eclipse, it was
ultimately considered a planet candidate but flagged as a possible
false positive. Finally, one vetter inspected each of the 87 flagged
candidates and issued a final disposition.

The number of candidates that survived this final vetting stage was
53. The candidates that were demoted included 1 which was a duplicate
of an accepted candidate, 19 which were deemed to be spurious
(i.e. systematic artifacts) or otherwise failing to have a consistent
shape and depth well above the photon noise, 2 which showed
out-of-transit variability in phase with the signal in question
(EPIC~212641218 and 212869892), and 12 which showed clear signs of
being an EB, a duplicate of an EB signal (i.e. half or double the
period), or having an ephemeris match to an EB. Finally, the
candidates from the K2-like vetting were subjected to further cuts
which are described in Sec.~\ref{sec:poe}.

Close inspection of the light curves of the planet candidates revealed
interesting information about a select number of candidates, which we
summarize below in Sec.~\ref{sec:indiv}.


\section{Stellar and Planetary Candidate Parameters}
\label{sec:params}
\begin{figure}[htb]
\begin{center}
\includegraphics[width=0.5\textwidth]{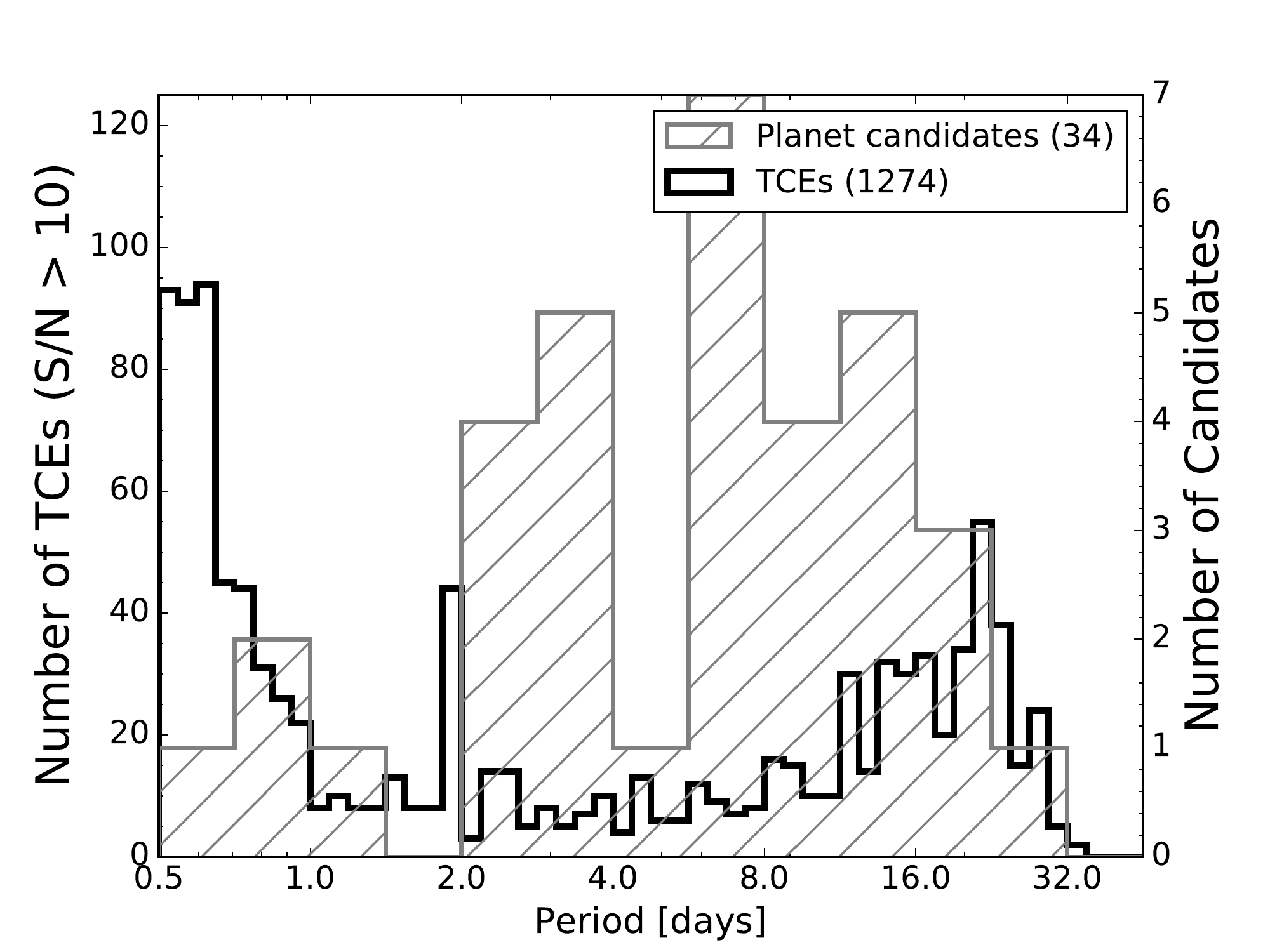}
\caption{\label{fig:per_dist} Orbital periods of planet candidates
  identified in our analysis. The dark, narrow-binned histogram
  (axis at left) shows the Threshold-Crossing Events (TCEs) identified
  by \TERRA with S/N$\ge$10 (see Sec.~\ref{sec:methods}).
  The gray, hatched histogram (axis at right) indicates the distribution of
  \npc planet candidates. }
\end{center}
\end{figure}

\begin{figure}[htb]
\begin{center}
\includegraphics[width=0.5\textwidth]{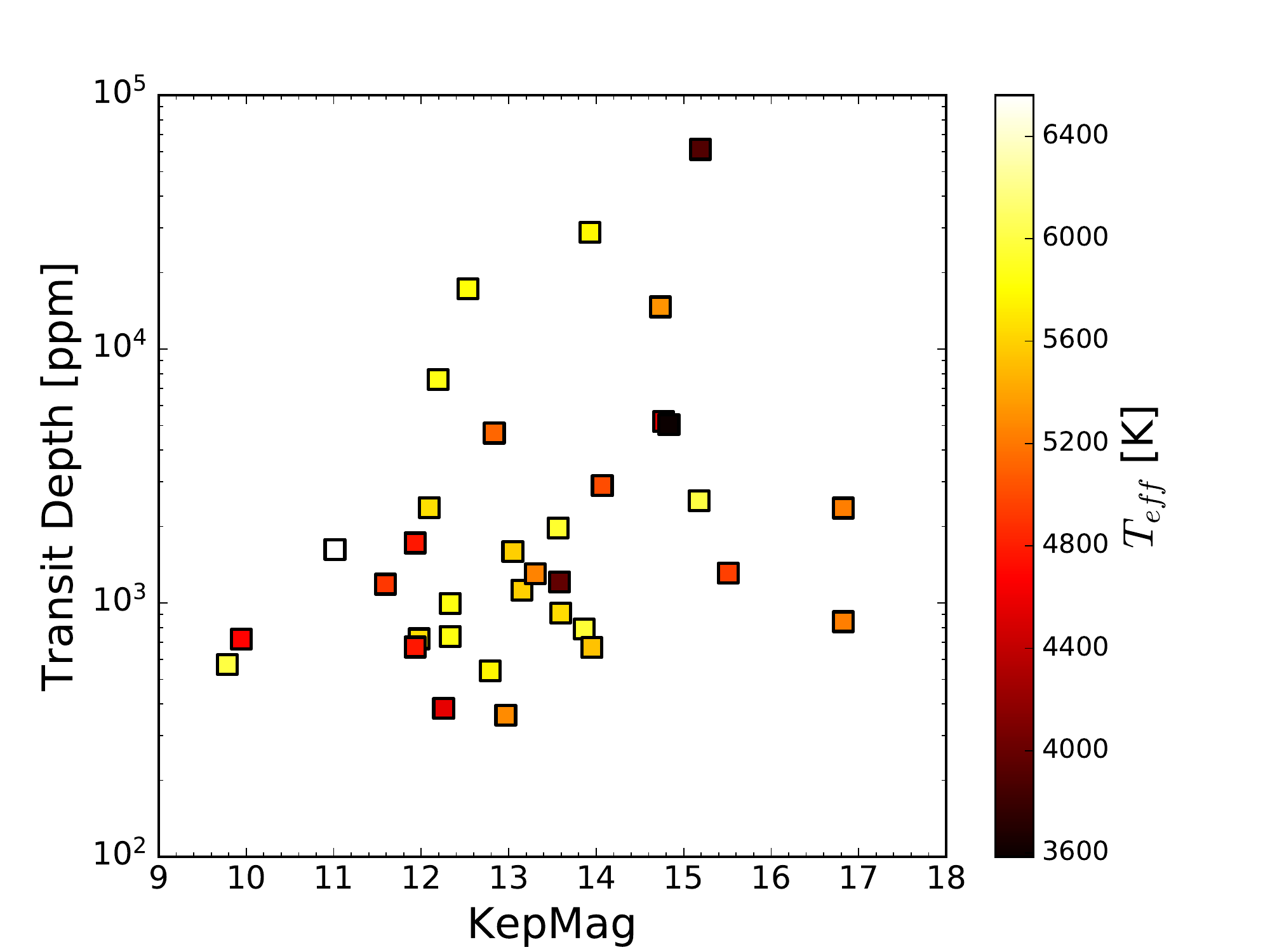}
\caption{\label{fig:kepmag-trandep} Transit depth and stellar
  magnitude for our planet candidates, as a function of stellar
  \teff (color scale). The two brightest targets are Wolf~503
  (EPIC~212779563) and HD~119130 (EPIC~212628254). }
\end{center}
\end{figure}

At the conclusion of the vetting exercises described above, we have
two lists of possible planet candidates with only a few physical
parameters known. Of these, the most salient are a candidate's orbital
period (shown in Fig.~\ref{fig:per_dist}) along with transit depth and
apparent stellar brightness (shown in Fig.~\ref{fig:kepmag-trandep}).
Stellar parameters for C17 stars are not available in the Ecliptic
Planet Input Catalog (EPIC) as they were in past K2 campaigns
\citep{huber:2016}, so the next step is to infer physical parameters
such as radii and temperatures.

\subsection{Ground-based Spectroscopy}
Happily, EPIC parameters and ground-based stellar spectroscopy exist
for some C17 stars also observed in C6. \cite{dressing:2017a} describe
medium-resolution infrared spectroscopy of late-type systems using
IRTF/SpeX, and \cite{petigura:2018} describe high-resolution optical
spectroscopy with Keck/HIRES of a broader sample.  Numerous spectra
have also been acquired with the Tillinghast Reflector Echelle
Spectrograph \citep[TRES;][]{furesz:2008} and uploaded to the
ExoFOP-K2 website; we describe these observations below.
Table~\ref{tab:spec} lists the key stellar parameters reported for 24
targets in C17 from SpeX, HIRES, and TRES.  We also include parameters
of two newly identified candidates orbiting bright stars from C17,
EPIC~212628254 and~212779563.

TRES is located on the 1.5-m Tillinghast Reflector at Fred Lawrence
Whipple Observatory on Mount Hopkins. TRES is a fiber-fed
cross-dispersed echelle spectrograph with a resolving power of $R
\approx 44,000$\ and an instrumental velocity precision of $10$\ to
$15$\,\ms, well-suited to stellar classification and identification of
binaries via radial velocity variations and/or composite spectra. We
use the Stellar Parameter Classification (SPC) package
\citep[see][]{buchhave:2012} to determine the effective temperature,
surface gravity, metallicity, and rotational broadening of each
spectrum, and we report those values in Table~\ref{tab:spec}. We also
report the radial velocities derived from the cross-correlation of a
single spectral order against the best-matched synthetic spectrum,
shifted to the absolute IAU scale. The TRES spectra---along with plots
of stellar classifications resulting from cross-correlation against a
coarse grid of synthetic spectra and spectral regions of
interest---are available on
ExoFOP-K2\footnote{\url{https://www.exofop.ipac.caltech.edu/k2}}.

\subsection{Multicolor Photometry and Gaia DR2}
Despite the spectroscopic data from SpeX, HIRES, and TRES, we desire a
complete and homogeneous set of stellar parameters against which to
compare our C17 candidate sample. To this end, we set aside
spectroscopic parameters and instead use EPIC multicolor
($BVugrizJHK$) photometry, parallaxes from Gaia DR2
\citep{gaia:2016a,gaia:2018}, and
\texttt{isochrones}\footnote{\url{https://github.com/timothydmorton/isochrones/}}
\citep{morton:2015b} to derive stellar parameters using the MIST
isochrones \citep{dotter:2016,choi:2016}.

\begin{figure}[htb]
\begin{center}
\includegraphics[width=0.5\textwidth]{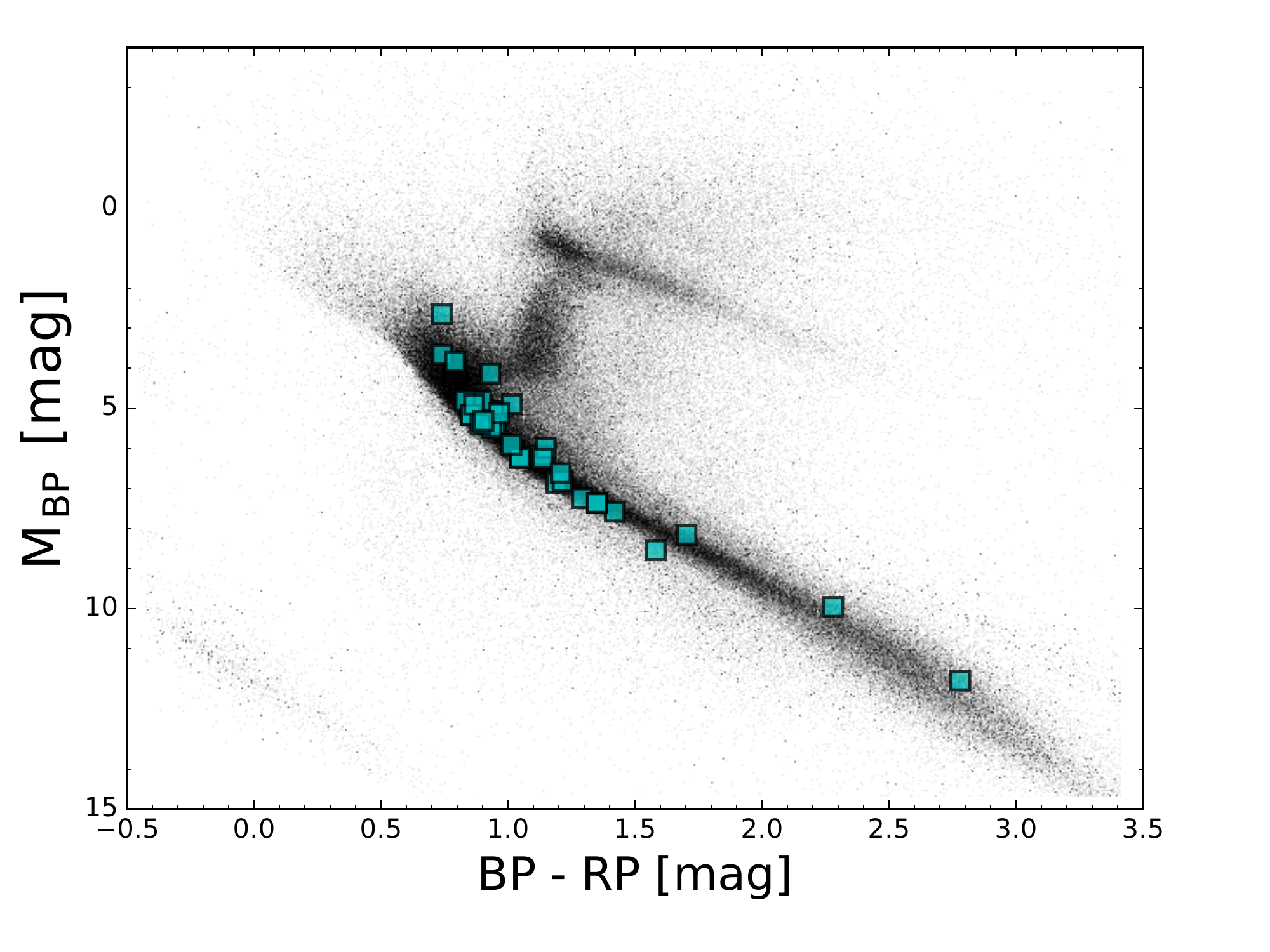}
\caption{\label{fig:cmd} Color-magnitude diagram for our C17 planet
  candidates (squares) and for all K2 targets (gray background).   }
\end{center}
\end{figure}

For C6 targets we use the Gaia-K2 cross-match from
\url{https://gaia-kepler.fun}. For targets not in C6 we run our own
cross-match between the EPIC locations and Gaia DR2 using an initial
search radius of 5'', selecting the Gaia source that most closely
matches the position and magnitude of the K2 target. There were no
ambiguous cases. All stars with $|Kp-G| > 0.5$ turned out to be stars
where $Kp$ was estimated from 2MASS colors alone.  For all planet
candidates, we are pleased to find that the distances inferred from
\texttt{isochrones} are consistent with those from Gaia (at the
$3\sigma$ level). The inferred stellar parameters for our candidates
are listed in Table~\ref{tab:c17} and are online at ExoFOP-K2, and a
color-magnitude diagram of our final candidate sample is shown in
Fig.~\ref{fig:cmd}.

\section{Results and Discussion}
\label{sec:results}

\subsection{Purifying the Sample}
\label{sec:poe}
Some of the TCEs that we identified as planet candidates subsequently
turned out to be non-planetary. Eleven candidates were identified as
planet candidates during TESS-like group vetting, but were
subsequently eliminated because the implied candidate radii would be
$>30R_\oplus$. These stars are EPIC~212579164, 212580081, 212627712,
212628098, 212770429, 212651213, 212757601, 212769367, 212769682,
212871068, and 212884586.

For the last of these, 212884586, Gaia DR2 shows two stars near the
source's location with $G$=19.8 and 19.6~mag, both located at
distances $>$400~pc and both within the K2 aperture.
Either could be the transit host and the transit would be diluted by
the light of the other, in which case our inferred radius of
$20^{+21}_{-13}R_\oplus$ would reach $\sim$30\,$R_\oplus$. We
therefore exclude this system from our planet candidate list.

We list EPIC~212658818 as an EB because its transit depth varies
throughout the campaign, both in C17 and in C6. This variation is
likely due to the putative transits occurring around a secondary star
12'' to the south that is partly in the K2 aperture. Ground-based
followup
photometry\footnote{\url{https://exofop.ipac.caltech.edu/k2/edit_target.php?id=212658818}}
indicates that this secondary star, fainter by 4.1~mag, is the true
host of the eclipses (which have a depth of 42\%).

We originally identified an EB and a planet candidate around
EPIC~212651213 and 251810686, but then discovered that both EPIC stars
target the same system (with an offset in the K2 data ``postage
stamp'' for EPIC~251810686).  We also acquired a light
curve\footnote{\url{https://exofop.ipac.caltech.edu/k2/edit_target.php?id=212651213}}
confirming an event depth of 9\% at our measured ephemeris. However,
we remove both systems from our candidate list because this is a
known quintuple system with two eclipsing binaries \citep{rappaport:2016}.

We note that several remaining candidates have radii formally below
our 30\,$R_\oplus$ limit, but are still grazing transits and so have
large radius uncertainties (e.g., 212628477 and 212686312). As
currently formulated, the TESS vetting process would report these as
candidates, so we retain them in our C17 sample with a note in
Table~\ref{tab:c17}.

\subsection{Planet Candidates, EBs, and Variables}

Our TESS-like vetting identified \npc\ planet candidates, all of which
were marked as candidates in K2-like vetting.  Our standard K2 vetting
process identified 53 planet candidates, but several of these were not
marked as candidates in TESS-like vetting for  reasons including:
\begin{itemize}
\item 251504891.01:  Marked as variable because of coherent out-of-transit variation.
\item 212473154.01:  Marked as EB because the candidate radius $R_C=65 R_\oplus$.
\item 212789681.01:  Marked as EB because the transit duration $T_{14}=0.12$~d is a large fraction of $P=0.49$~d.
\item 212421319.01:  Marked as EB because the odd and even transits have different depths.
\item 212499716.01:  Marked as EB because of a faint secondary eclipse, seen more clearly in C6 photometry.
\item 212579164.01:  Marked as EB because $R_C=46 R_\oplus$.
\item 212580081.01:  Marked as EB because $R_C=35 R_\oplus$.
\item 212627712.01:   Marked as IS because  the K2 photometric aperture mostly captures light from a nearby, brighter star.
\item 229228115.01:   Marked as EB because $T_{14}=0.13$~d is a large fraction of $P=0.55$~d.
\item 212705192.01: Marked as EB because of odd-even effect, and because Keck/HIRES
  and TRES spectra show the star to be double-lined.
\item 212740148.01:  Marked as EB because of a faint secondary eclipse. Also, the K2 photometric aperture mostly captures light from a nearby, brighter star.
\item 212770429.01:  Marked as IS because the K2 photometric aperture mostly captures light from a nearby, brighter star. 
\end{itemize}



\begin{figure*}[htb]
\begin{center}
\includegraphics[width=\textwidth]{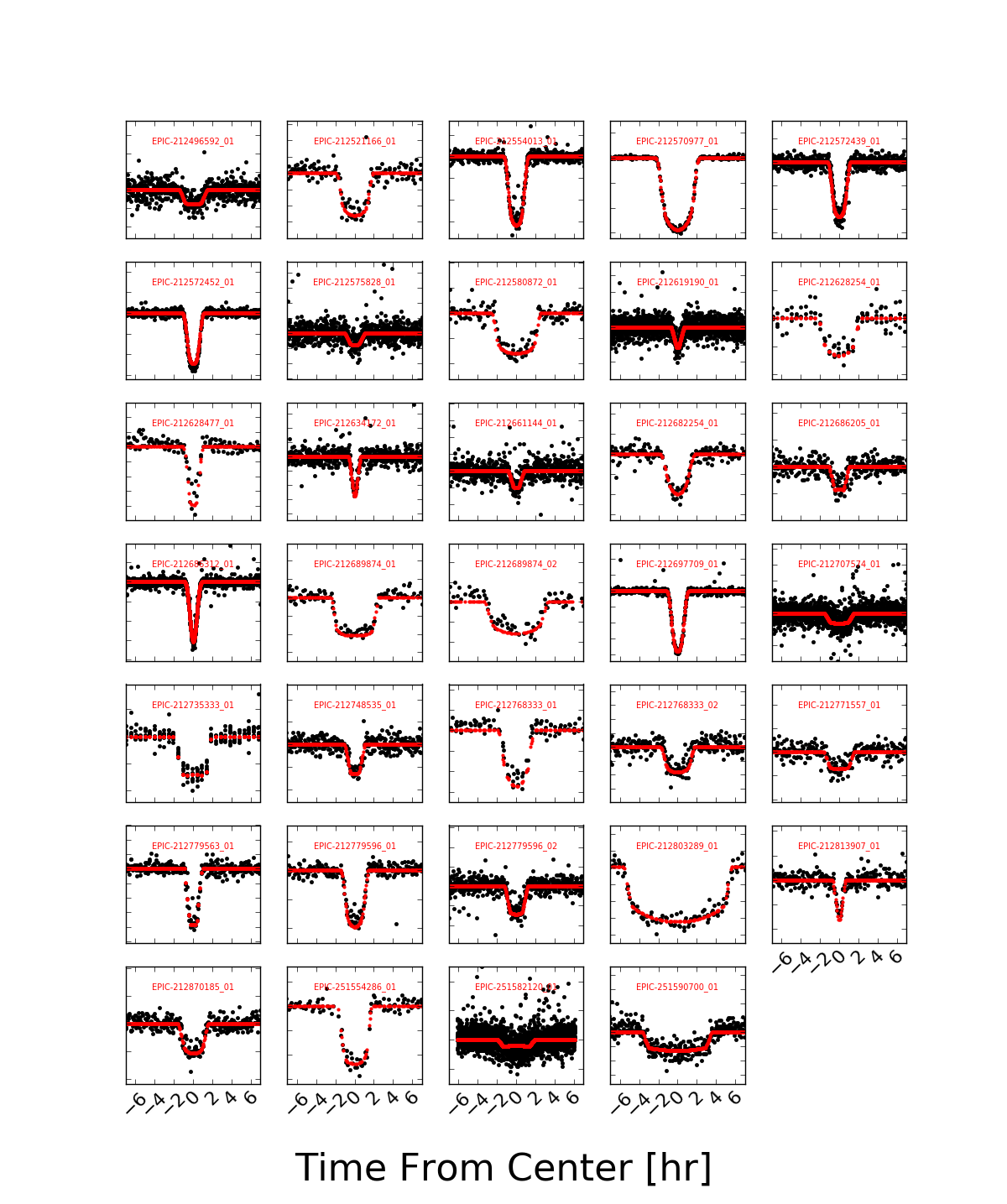}
\caption{\label{fig:hi_cand} Phase-folded light curves of our
  \npc\ planet candidates, and their best-fit transit models. To show
  all transits, the vertical scale is different in each panel; system
  parameters are listed in Table~\ref{tab:c17}.  }
\end{center}
\end{figure*}

\begin{figure}[htb]
\begin{center}
\includegraphics[width=0.5\textwidth]{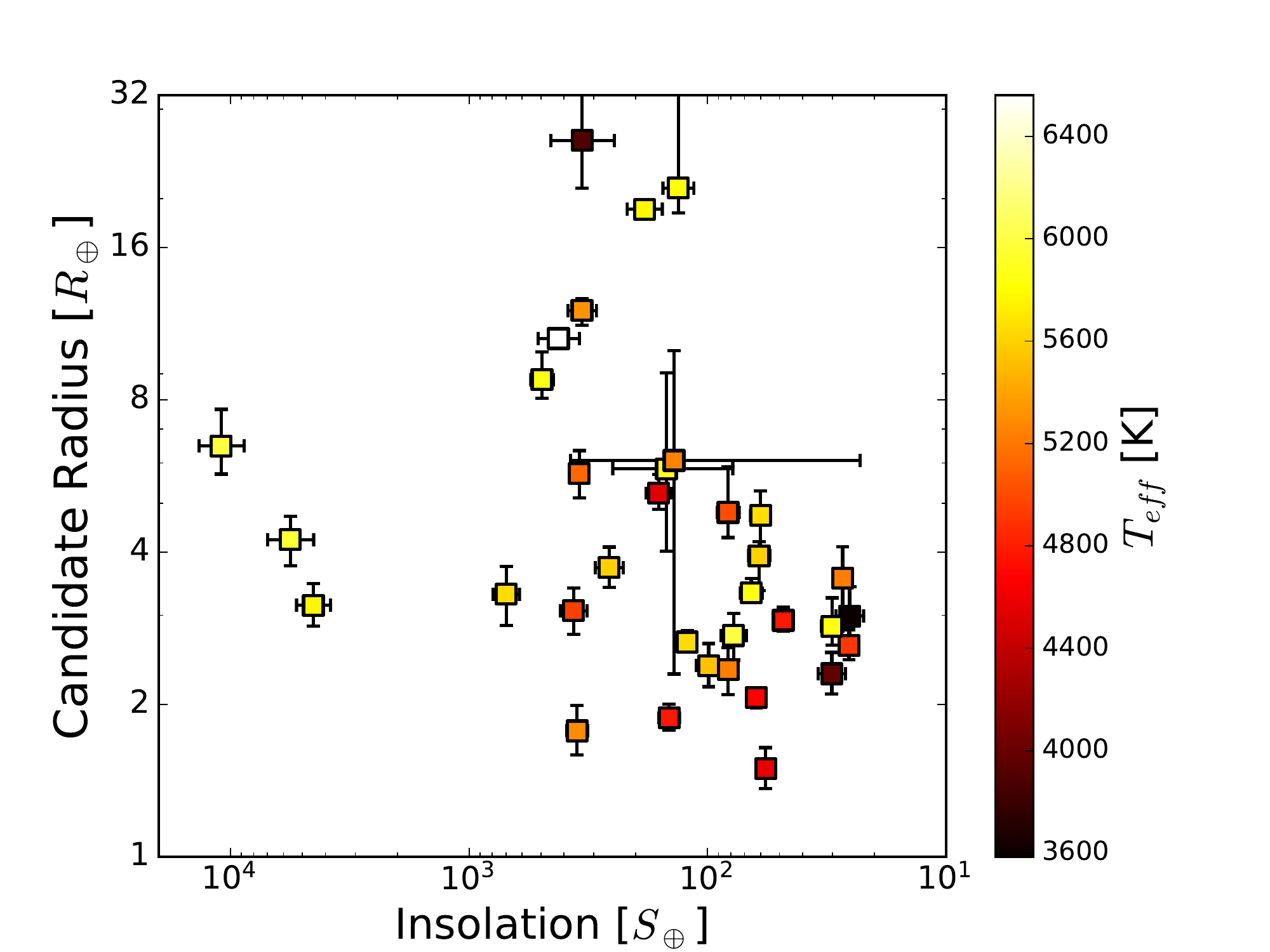}
\caption{\label{fig:insol-rad} Candidate radius and incident
  insolation for our planet candidates, as a function of stellar
  \teff (color scale).}
\end{center}
\end{figure}

Table~\ref{tab:c17} lists the basic parameters for our final list of
\npc\ planet candidates from K2's C17. The properties of this
population are also summarized in Fig.~\ref{fig:per_dist} (orbital
periods), Fig.~\ref{fig:hi_cand} (phase-folded candidate light
curves), Fig.~\ref{fig:kepmag-trandep} ($Kp$ and transit depth),
and Fig.~\ref{fig:insol-rad} (candidate radius and insolation).

We also include a list of all likely EBs and other apparently
astrophysical variables identified from our TESS-like analysis. A
total of \neb\ EBs are listed in Table~\ref{tab:eb}, and
\nvar\ variables are listed in Table~\ref{tab:var}.  These tables also
include the final comments (if any) assigned to each TCE during the
group vetting process. Note also that the numbers above likely
somewhat overestimate the objects in each category, since EBs with
secondary eclipses and variables with multiple harmonics are both
often identified as multiple TCEs in the same system.

\section{Comparing Planet Candidates: C17 vs.\ C6}
\label{sec:comparison}
Twenty-one of our planet candidates (orbiting 18 stars) were also
observed by K2 in C6. This earlier campaign was searched for
transiting planets by many groups, giving us a rare opportunity to
compare the results of these analyses.  Different teams have used a
variety of photometric and transit search pipelines, all using fully
calibrated data products. Because our analysis here uses raw cadence
data (calibrated only by \texttt{kadenza}), our noise levels are
higher and we do not expect to identify all transit-like signals
described in the literature. Although we might naively expect
substantial or complete overlap between the C6 surveys, that is not
what we find. Table~\ref{tab:c6} compares the disposition of these 21
C6+C17 candidates by several large-scale surveys, which we describe
below.

\cite{pope:2016} identify 19 of our candidates as planet candidates,
missing only two of our candidate systems --- EPIC~212634172 and
212686205. This is the highest degree of overlap for any C6 catalog,
suggesting a higher completeness rate than  other analyses.

\cite{dressing:2017a,2017b} derive stellar and planetary parameters and
associated false positive probabilities for planets orbiting late-type stars that were discovered by 
multiple transit surveys. They
validate EPIC~212554013 and 212686205, leave 212634172 as a
planet candidate, and deem 212572452 to be a false positive
because its  photometry is blended with that of 212572439.

\cite{mayo:2018} identify and validate planets in ten of our
candidate systems: EPIC~212496592, 212521166, 212580872, 212686205,
212689874, 212697709, 212735333, 212768333, 212779596, and 212803289.
They do not report any candidates around our candidate systems
EPIC~212554013, 212570977, 212572452, 212572439, 212575828, 212634172,
212661144, or 212813907.

Finally, the signals in 11 of our C6+C17 systems were identified as planet
candidates by \cite{petigura:2018}, {\em viz.}, EPIC~212521166,
212554013, 212570977, 212572452, 212572439, 212580872, 212689874,
212697709, 212735333, 212779596, and 212803289.  In a follow-up paper,
Livingston et al.\ (submitted) validate EPIC~212521166, 212554013,
212580872, 212689874, and 212779596.  EPIC~212697709 remains a
candidate in the latter paper with a false positive probability of
1.9\%, but this planet was validated as WASP-157 \citep{mocnik:2016}.
Livingston et al.\ also find a sufficiently low FPP to validate
EPIC~212803289 and 212570977, but out of an abundance of caution they
deem these to be candidates because of their large radii
($>10R_\oplus$).  They also find EPIC~212572439 and 2127355333 to 
have very low FPPs but call these merely candidates because of an
additional stellar source in the K2 photometric aperture (Gonzales et
al., in prep.).

As a further comparison, we calculated the ephemerides offsets of
eleven of our C17 candidates with those derived from C6 data.  To
avoid possible biases that could arise from using different pipelines,
we only compared those candidates with ephemerides reported by
Livingston et al. (submitted). Ephemerides for all eleven candidates
are consistent at the 3$\sigma$ level, with only three candidates
disagreeing at the 2--3$\sigma$ level (212570977.01, 212779596.01, and
212803289.01).

\begin{deluxetable*}{l c c c c c l l}[bt]
\tabletypesize{\scriptsize}
\tablecaption{Our C17 Candidates Observed in C6 \label{tab:c6}}
\tablehead{
\colhead{Candidate } & \colhead{ C6 } & \colhead{ Po16 } & \colhead{ Ma18 } & \colhead{ Pe18 } & \colhead{ Li18 } & \colhead{ Name } & \colhead{ Validation Reference / Note}
}
\startdata
212496592.01 & Y & PC & VP & N  & N  & K2-191b &  \cite{mayo:2018}\\
212521166.01 & Y & PC & VP & PC & VP & K2-110b &   \cite{osborn:2017}\\
212554013.01 & Y & PC & N  & PC & VP & K2-127b &   \cite{dressing:2017b}\\
212570977.01 & Y & PC & N  & PC & PC & --- &  ---    \\
212572439.01 & Y & PC & N  & PC & PC & --- &  Blend with 212572452.\\ 
212572452.01 & Y & PC & N  & N  & PC & --- &  Blend with 212572439. \\ 
212575828.01 & Y & PC & N  & N  & N  & --- &  ---    \\
212580872.01 & Y & PC & VP & PC & VP & K2-193 &  \cite{mayo:2018}\\
212634172.01 & Y & N  & N  & N  & N  & --- &  --- \\
212661144.01 & Y & PC & N  & N  & N  & --- &  ---    \\
212686205.01 & Y & N  & VP & N  & N  & K2-128b &  \cite{dressing:2017b}\\
212689874.01 & Y & PC & VP & PC & VP & K2-195b &  \cite{mayo:2018}\\ 
212689874.02 & Y & PC & VP & PC & VP & K2-195c &   \cite{mayo:2018}\\
212697709.01 & Y & PC & VP & PC & PC & WASP-157b &    \cite{mocnik:2016} \\ 
212735333.01 & Y & PC  & VP  & PC  & PC & K2-197b & \cite{mayo:2018}\\ 
212768333.01 & Y & PC & VP & N  & N  & K2-198b  &   \cite{mayo:2018}\\ 
212768333.02 & Y & PC & N & N  & N  &  &   --- \\
212779596.01 & Y & PC & VP & PC & VP & K2-199b &  \cite{mayo:2018}\\ 
212779596.02 & Y & PC & VP & PC & VP & K2-199c &  \cite{mayo:2018}\\ 
212803289.01 & Y & PC & VP & PC & PC & K2-99b &  \cite{smith:2017} \\ 
212813907.01 & Y & PC & N  & N  & N  &  --- &  ---    \\
\enddata
\tablenotetext{}{References: Po16 \citep{pope:2016}, Ma18 \citep{mayo:2018}, Pe18 \citep{petigura:2018}, Li18 (Livingston et al., submitted).}
\tablenotetext{}{Notes: VP (validated planet), PC (planet candidate), N (not identified).}
\end{deluxetable*}






\section{Individual Systems}
\label{sec:indiv}
Below we discuss several interesting individual systems discovered by
our C17 analysis. We separate these into several groups: potentially
exciting discoveries warranting additional follow-up observations;
more generic candidates nonetheless requiring some additional
discussion; and finally, objects which (though planet candidates) may
be somewhat more likely to be non-planetary false positives.

\begin{itemize}
\item 212779563 (Wolf 503, HIP 67285). This candidate planet's size
  of 2$R_\oplus$ lies near the gap between sub-Neptunes and
  super-Earths \citep{fulton:2017}. The short period and nearby,
  bright star (V=10.3, H=7.8) could make this an excellent target for
  future RV and transmission spectroscopy. This system is described in
  more detail by Peterson et al.\ (submitted).

\item 212628254 (HD 119130). This 2.7\,$R_\oplus$ candidate orbits
  a V=9.9, slightly evolved G star. It may also be a good RV target
  because of the planet's moderate size and bright host star.

\item 212813907: In addition to the transiting planet candidate
  reported here with $P=6.7$~d, we see an obvious single transit with
  depth 1.8\% centered at BJD$_{TBD}$=2458213.82646 and with
  duration 0.66~d.  This points to a candidate transiting companion
  with a radius of $\sim 1 R_\mathrm{Jup}$ and $P\approx1000$~d. No
  corresponding transit was seen for this star during C6.

\item 212686205 (K2-128).  \citep{dressing:2017a} showed that this
  star is a K4 dwarf, despite its EPIC classification as a giant
  \citep{huber:2016}.  The star exhibits semi-sinusoidal brightness
  variations that are likely due to starspots and stellar surface
  rotation, with a period of $P_\mathrm{rot}$=11.9 days and amplitude
  of 0.018 mag. The position of the star in a rotation period-color
  diagram indicates an age similar to that of Praesepe ($\sim$600--800
  Myr).

\item 212768333: This candidate was validated as the single-planet
  K2-198 b ($P=17$~d) using data from C6 \citep{mayo:2018}, but our
  C17 data also reveal a second candidate with $P=7.4$~d. These two
  candidates, plus a third ($P=3.4$~d) were previously reported by
  \cite{pope:2016}. The star has K2 data available from Campaigns 6
  and 17, making a search for additional transiting planets at longer
  orbital periods possible. The star shows periodic variability which
  is likely due to rotation of the spotted surface. The inferred
  rotation period of 7.02 days and variability amplitude of 0.024 mag
  (from the 10th to 90th percentile) point to a young system age
  \citep{rebull:2016,rebull:2018}, likely older than the Pleiades
  (125~Myr) but perhaps younger than or similar in age to Praesepe
  ($\sim$600--800 Myr).

\item 212619190 and 212707574: These are both ultra-short-period (USP)
  planet candidates. While the signals are convincing, the inferred
  sizes we report here are larger than typical USPs
  \citep{winn:2018}. 

\end{itemize}
\noindent The following planet candidates seem reliable but warrant some additional discussion.

\begin{itemize}

\item 212748535 -- We originally identified this candidate as a signal
  associated with EPIC~212748598 ($Kp$=17.4~mag).  This faint source
  is classified as a galaxy by The 2dF Galaxy Redshift Survey
  \citep{colless:2001} and appears galaxy-like in Pan-Starrs
  multicolor imaging (A.~Rest, private communication). We conclude
  that EPIC~212748598 is a galaxy despite its designation as ``STAR''
  in EPIC.  Gaia DR2 shows a brighter, stellar source with $\Delta
  G=5.4$~mag within our K2 aperture and 20'' away. This brighter star
  is EPIC~212748535, which Gaia shows to be a K dwarf
  (\teff=3800~K, $R_*=0.67 R_\odot$) and which dominates the flux
  in our K2 photometric aperture. We conclude that the brighter
  source, EPIC~212748535, is the true host of the observed
  $\sim$1~mmag transit.

\item 212682254: This star has a candidate with $R_C=6R_\oplus$ and
  $P=10.7$~d, and also shows photometric variability due to starspots,
  with an amplitude of 0.019 mag (again measured from the 10th to 90th
  percentile) and an inferred rotation period of 9.45 days.  The
  rotation period and color place the star near the slowly-rotating
  I-sequence of Praesepe members \citep{barnes:2007gyro}, indicating
  an age similar to that cluster ($\sim$600--800 Myr).

\item 212572439 and 212572452: Our analysis independently identified
  two candidates with the same periods around these adjacent stars
  (separated by 6''). A transit-like signal from the blend of these
  two sources has also been identified in previous works
  \citep[][Livingston et al., submitted; Gonzales et al., in
    prep.]{dressing:2017b,petigura:2018}, and both signals were
  identified (though the blend went unremarked) by \cite{pope:2016}.
  Based on our inferred stellar and planetary properties, this signal
  could still be a transiting planet regardless of which of these two
  stars it orbits; we thus retain both signals as planet
  candidates. Additional follow-up will be required to identify which
  object is the transit host.

\end{itemize}
\noindent Finally, the objects below pass our criteria as planet candidates but show warning signs hinting that they may be non-planetary:

\begin{itemize}
\item 251590700: This source has no Gaia DR2 parallax so the derived
  stellar parameters are somewhat less certain. The parallax
  measurement is presumably lacking because of an enormous amount of
  excess noise in the five-parameter Gaia solution
  (\texttt{astrometric\_excess\_noise\_sig}=64781), suggesting the
  possibility that the star is a binary. Our transit fit implies a
  stellar density \citep[assuming a circular orbit;][]{seager:2003} of
  $\rho_{*,circ}=0.0033^{+0.0005}_{-0.0003}$~g~cm$^{-1}$, implying
  either a highly eccentric orbit or a false positive caused by an
  eclipsed, low-density giant star.

\item 251582120: We originally identified this event as a signal
  around EPIC~251581990, a faint ($Kp$=18.5~mag) source listed as an
  ``EXTENDED'' (i.e., non-stellar) object in EPIC. Our aperture for
  this faint target enclosed another nearby brighter stellar source,
  EPIC~251582120 ($Kp$=15.2~mag), whose flux dominates  our light
  curve. Our light curve fit for this brighter source implies
  $\rho_{*,circ}=0.165\pm0.055$~g~cm$^{-1}$, mildly inconsistent with our
  \texttt{isochrones}+Gaia-derived stellar density of
  $0.79\pm0.20$~g~cm$^{-1}$. The crowded aperture and mismatch in
  stellar densities hint that this planet candidate may be less
  reliable.


\item 212686312: This signal is both deep and V-shaped, indicating a
  grazing transit. Combined with the very short orbital period and the
  inferred companion radius presented here, the planetary nature of
  the signal is doubtful.

\item 212628477: This star is rapidly rotating, with a period of 2.685
  days and a variability amplitude of 0.045 mag. The star's rapid
  rotation combined with its color suggest an age younger than that of
  the Pleiades \citep{rebull:2016}. The rotation period is clearly
  distinct from the much longer period of the planet candidate
  ($P$=15.4~d), but there are several warning signs for this
  candidate: the transits are grazing so the inferred companion is
  large ($21.0^{+15.4}_{-2.2} R_\oplus$); Gaia DR2 reports a highly
  uncertain radial velocity of 20.98$\pm$19.55~km~s$^{-1}$, perhaps
  indicative of RV variability; and the TRES spectrum shows a probable
  shoulder in the cross-correlation function indicating a
  double-lined spectrum (see Table~\ref{tab:spec}).

\item 251539584 and 251539609: These two stars are both spectroscopic
  binaries.  Both showed candidate transit signals with the same
  transit ephemeris ($P=1.09$~d). The stars are roughly equal
  brightness ($\Delta Kp$=0.2~mag) and are separated by roughly 14''
  and are both are contained in the photometric aperture applied to
  the other. The two stars are apparently associated and co-moving,
  based on their kinematics from Gaia DR2. The combined light curve is
  variable, indicating a rotation period of 4.34 days and amplitude of
  0.002 mag (though the true amplitude must be larger because of flux
  dilution from the companion). TRES spectroscopy shows that both EPIC
  sources are short-period double-lined spectroscopic binaries (see
  Table~\ref{tab:spec}), so we list these systems as candidate EBs.






\end{itemize}

\section{Discussion and Conclusion}
\label{sec:conclusions}

From \nstars\ stars observed in K2's most recent field, Campaign 17,
we identified \ntce\ transit-like events. Among these, we find
\npc\ planet candidates (Table~\ref{tab:c17}), \neb\ eclipsing
binaries (Table~\ref{tab:eb}), and \nvar\ other periodic variables
(Table~\ref{tab:var}). Because C17 was observed in ``forward-facing''
mode by K2 in its Earth-trailing orbit, these targets can be
immediately observed before the ecliptic field sets for the
season. Many of these objects were also observed by K2 during C6,
offering a rare opportunity to study the same systems over a 1000~day
timespan. Multiple observations of the same field will be commonplace
when TESS begins near-continuous observations of the ecliptic poles,
which will substantially increase that survey's sensitivity to
long-period planets. Though beyond the scope of this work, a
comprehensive transit search in C6+C17 (or C5+C16) would probe a
single, narrow range of orbital periods from 880--1030~d (and
harmonics of these periods).

We evaluated the overlap between our C17 planet candidates and those
observed in C6 by several earlier planet surveys, finding again that
K2 efforts have substantially different completeness
\citep{crossfield:2016,mayo:2018}. The C6 catalog of \cite{pope:2016}
overlaps most closely with our C17 candidate list, indicating that
that sample has either a high degree of completeness or (at worst) a
very similar set of biases to that of our sample. Unfortunately,
the different samples and data quality between the calibrated C6 data
and our use of C17's raw cadence data precludes any conclusions about
false positive rates in these surveys. Nonetheless, the generally
incomplete overlap between the candidate lists of different surveys
lends support to the TESS science plan to use two independent
pipelines, SPOC and QLP, to minimize the chances of interesting planet
candidates passing unnoticed.

In this work we focus on the search for new transiting planet
candidates, whose parameters are summarized in Table~\ref{tab:c17}. We
find several candidates that have sizes $<$4$R_\oplus$ and orbit stars
with $Kp\lesssim10$, indicating that these are good RV targets. The
most interesting are Wolf 503 (EPIC~212779563.01; see Peterson et al.,
submitted) and HD~119130 (EPIC~212628254.01). If found by TESS, such
planet candidates would be ideal targets for fulfilling its prime
science goal of contributing to the measured masses of 50 small
planets.

Several other planet candidate discoveries highlight potentially
intriguing dynamical and/or multi-body systems.   We see
 a single, deep transit around EPIC~212813907, which also hosts a
6~d planet candidate, suggesting a Jupiter-sized companion on a
long-period orbit.  We also identify a candidate planet in each of two
possible binary systems (EPIC~251539584 \& 251539609, and
EPIC~212572439 \& 212572452).


In conclusion, K2's rapid data releases for its recent campaigns have
facilitated quick identification of many interesting astrophysical
phenomena in time for immediate ground-based follow-up. This approach
is qualitatively the same as that planned for TESS.  In this C17
exercise, our TESS-like and K2-like vetting approaches both yielded
the same set of planet candidates. This result validates the results
derived from similar, past analyses of K2 and also demonstrates that
the team members soon to be examining TESS data have the tools and
expertise necessary for a successful mission.  After four years {\em
  Kepler} yielded to {\em K2}; another four years on, in Olympic
fashion {\em K2} will likewise pass the baton to {\em TESS} to
continue building on the great legacy of exoplanet exploration.

\acknowledgments

We thank A.~Rest for discussions about the nature of EPIC~212748598.

I.J.M.C.\ acknowledges support from NASA through  K2GO grant 80NSSC18K0308 and from NSF through grant AST-1824644.

This paper includes data collected by the \kep\ mission. Funding for the \kep\ mission is provided by the NASA Science Mission directorate. Some of the data presented in this paper were obtained from the Mikulski Archive for Space Telescopes (MAST). STScI is operated by the Association of Universities for Research in Astronomy, Inc., under NASA contract NAS5--26555. Support for MAST for non--HST data is provided by the NASA Office of Space Science via grant NNX13AC07G and by other grants and contracts. This research has made use of the Exoplanet Follow-up Observing Program (ExoFOP), which is operated by the California Institute of Technology, under contract with the National Aeronautics
and Space Administration.

Facilities: \facility{Kepler}, \facility{K2}, \facility{FLWO:1.5m
  (TRES)}, \facility{KeckI (HIRES)}, \facility{APF (Levy)}


\bibliographystyle{apj_hyperref}

\begin{turnpage}
\begin{deluxetable*}{lrlllllllll}[bt]
\tabletypesize{\footnotesize}
\tablecaption{Planet Candidates From C17 \label{tab:c17}}
\tablehead{
  {} & {Kp} & { $P$ } & {$T_0$ } & { $T_{14}$ } & { $Rp/R*$ } & { $R_*$ } & { $T_{eff}$ } & {   $R_P$ } & { $S_{inc}$ } & { Notes} \\
  {Candidate } & {  [mag] } & { [d] } & {BJD$_{TDB}$ - 2454833 } & { [hr]} & { [\%] } & { [$R_\odot$] } & { [K] } & {   [$R_\oplus$] } & { [$S_{\oplus}$] } & {}
}
\startdata
 212496592.01 &  12.966 &         $2.85883^{+0.00039}_{-0.00038}$ &      $3347.0222^{+0.0047}_{-0.0053}$ &       $2.17^{+0.40}_{-0.29}$ &      $1.89^{+0.23}_{-0.20}$ &  0.86 &      5284 &      $1.77^{+0.22}_{-0.19}$ &          352               &   K2-191b \citep{mayo:2018}     \\                                                                        
 212521166.01 &  11.590 &           $13.8642^{+0.0011}_{-0.0011}$ &      $3357.3269^{+0.0028}_{-0.0027}$ &       $3.26^{+0.24}_{-0.18}$ &      $3.35^{+0.25}_{-0.21}$ &  0.72 &      4915 &      $2.62^{+0.20}_{-0.16}$ &       25.5                 &   K2-110b  \citep{osborn:2017}    \\                                                                   
 212554013.01 &  14.733 &      $3.588223^{+0.000046}_{-0.000045}$ &   $3348.97026^{+0.00046}_{-0.00047}$ &    $2.137^{+0.086}_{-0.073}$ &     $11.61^{+0.47}_{-0.70}$ &  0.95 &      5324 &     $12.01^{+0.65}_{-0.77}$ &          336               &     K2-127b \citep{dressing:2017b}  \\                                                                    
 212570977.01 &  13.928 &      $8.853181^{+0.000052}_{-0.000051}$ &   $3347.02423^{+0.00021}_{-0.00022}$ &    $4.192^{+0.029}_{-0.027}$ &     $15.33^{+0.22}_{-0.15}$ &  1.14 &      5774 &     $19.04^{+0.63}_{-0.62}$ &          183               &     \nd  \\                                                                                               
 212572439.01 &  12.835 &      $2.581446^{+0.000038}_{-0.000038}$ &   $3347.75306^{+0.00055}_{-0.00054}$ &       $1.81^{+0.23}_{-0.12}$ &      $6.17^{+0.67}_{-0.65}$ &  0.85 &      5124 &      $5.72^{+0.63}_{-0.60}$ &          344               &    Blend with 212572452 \\ 
 212572452.01 &  14.769 &      $2.581446^{+0.000019}_{-0.000020}$ &   $3347.75323^{+0.00030}_{-0.00028}$ &    $1.761^{+0.036}_{-0.039}$ &      $7.19^{+0.61}_{-0.50}$ &  0.67 &      4535 &      $5.23^{+0.46}_{-0.38}$ &          160               &     Blend with 212572439 \\ 
 212575828.01 &  15.508 &         $2.06033^{+0.00018}_{-0.00018}$ &      $3347.0331^{+0.0033}_{-0.0033}$ &       $1.55^{+0.27}_{-0.14}$ &      $3.71^{+0.38}_{-0.37}$ &  0.76 &      4949 &      $3.07^{+0.33}_{-0.32}$ &          364               &   \nd    \\                                                                                               
 212580872.01 &  13.047 &           $14.7881^{+0.0013}_{-0.0012}$ &      $3352.4604^{+0.0029}_{-0.0029}$ &       $4.34^{+0.74}_{-0.20}$ &      $3.70^{+0.24}_{-0.54}$ &  0.98 &      5586 &      $3.93^{+0.26}_{-0.57}$ &       60.8                 &  K2-193b \citep{mayo:2018}\\ 
 212619190.01 &  12.788 &      $0.911861^{+0.000032}_{-0.000036}$ &      $3347.2783^{+0.0015}_{-0.0013}$ &    $0.772^{+0.121}_{-0.069}$ &      $2.33^{+0.23}_{-0.20}$ &  1.23 &      5765 &      $3.14^{+0.33}_{-0.29}$ &       4494                 &     HD~119130  \\                                                                                            
 212628254.01 &   9.782 &           $16.9813^{+0.0022}_{-0.0022}$ &      $3347.2910^{+0.0044}_{-0.0046}$ &       $3.69^{+0.59}_{-0.31}$ &      $2.32^{+0.24}_{-0.24}$ &  1.08 &      5998 &      $2.74^{+0.29}_{-0.29}$ &      77.9                  &     ---  \\                                                                                           
 212628477.01 &  12.533 &        $15.42404^{+0.00081}_{-0.00097}$ &      $3347.7248^{+0.0020}_{-0.0019}$ &       $1.54^{+0.26}_{-0.23}$ &       $13.8^{+10.2}_{-1.4}$ &  1.39 &      5823 &       $21.0^{+15.4}_{-2.2}$ &          132               &     Grazing transit  \\                                                                                               
 212634172.01 &  14.831 &      $2.851770^{+0.000083}_{-0.000092}$ &      $3348.4657^{+0.0013}_{-0.0011}$ &    $0.721^{+0.140}_{-0.062}$ &      $7.27^{+0.98}_{-0.64}$ &  0.38 &      3585 &      $2.99^{+0.42}_{-0.30}$ &       25.4                 &      \nd \\ 
 212661144.01 &  13.595 &         $2.45875^{+0.00022}_{-0.00019}$ &      $3347.2747^{+0.0028}_{-0.0031}$ &       $1.10^{+0.29}_{-0.18}$ &      $3.10^{+0.41}_{-0.41}$ &  0.98 &      5647 &      $3.30^{+0.45}_{-0.44}$ &          698               &    \nd \\ 
 212682254.01 &  13.565 &        $10.70070^{+0.00088}_{-0.00090}$ &      $3353.1746^{+0.0027}_{-0.0028}$ &       $3.23^{+0.31}_{-0.34}$ &      $4.74^{+2.05}_{-0.93}$ &  1.12 &      5936 &         $5.8^{+3.2}_{-1.8}$ &         148                &     ---  \\                                                                                              
 212686205.01 &  12.256 &         $5.67623^{+0.00042}_{-0.00056}$ &      $3347.6471^{+0.0044}_{-0.0031}$ &       $1.45^{+0.21}_{-0.12}$ &      $2.05^{+0.20}_{-0.18}$ &  0.67 &      4566 &      $1.49^{+0.15}_{-0.13}$ &       57.1                 & K2-128b \citep{dressing:2017b} \\ 
 212686312.01 &  15.192 &   $0.7476280^{+0.0000027}_{-0.0000027}$ &   $3346.76330^{+0.00015}_{-0.00014}$ &    $1.434^{+0.079}_{-0.067}$ &       $45.4^{+10.8}_{-8.1}$ &  0.53 &      3904 &        $26.0^{+6.8}_{-5.1}$ &         335                &     Grazing transit  \\                                                                                 
 212689874.01 &  12.330 &           $15.8537^{+0.0013}_{-0.0013}$ &      $3359.2217^{+0.0024}_{-0.0023}$ &       $4.52^{+0.21}_{-0.15}$ &      $3.11^{+0.21}_{-0.12}$ &  0.98 &      5842 &      $3.32^{+0.23}_{-0.14}$ &       65.7                 &     K2-195b \citep{mayo:2018} \\                                                                       
 212689874.02 &  12.330 &           $28.4545^{+0.0034}_{-0.0034}$ &      $3349.1480^{+0.0044}_{-0.0041}$ &       $6.08^{+0.54}_{-0.40}$ &      $2.67^{+0.37}_{-0.21}$ &  0.98 &      5842 &      $2.85^{+0.39}_{-0.23}$ &       30.1                 &      K2-195c \citep{mayo:2018}\\ 
 212697709.01 &  12.193 &      $3.951632^{+0.000030}_{-0.000030}$ &   $3349.48035^{+0.00029}_{-0.00029}$ &       $1.82^{+0.12}_{-0.10}$ &      $7.40^{+1.01}_{-0.57}$ &  1.09 &      5860 &      $8.77^{+1.18}_{-0.71}$ &          494               &     WASP-157, K2-41 \citep{mocnik:2016} \\                                                                
 212707574.01 &  13.861 &         $1.12665^{+0.00018}_{-0.00014}$ &      $3346.9600^{+0.0047}_{-0.0067}$ &       $2.36^{+0.46}_{-0.28}$ &      $2.38^{+0.22}_{-0.25}$ &  1.63 &      5967 &      $4.24^{+0.48}_{-0.47}$ &     5618                   &    ---   \\                                                                                          
 212735333.01 &  11.977 &         $8.35812^{+0.00039}_{-0.00043}$ &      $3354.6901^{+0.0019}_{-0.0018}$ &       $3.30^{+0.16}_{-0.13}$ &      $2.63^{+0.13}_{-0.11}$ &  0.93 &      5642 &      $2.66^{+0.14}_{-0.12}$ &     121.8                  &    K2-197b \citep{mayo:2018} \\                                                                      
 212748535.01 &  13.582 &         $5.47826^{+0.00034}_{-0.00033}$ &      $3349.3152^{+0.0021}_{-0.0020}$ &       $1.53^{+0.21}_{-0.15}$ &      $3.51^{+0.33}_{-0.29}$ &  0.60 &      3971 &      $2.30^{+0.23}_{-0.20}$ &       30.2                 &    \nd \\                                                                                              
 212768333.01 &  16.825 &        $17.04518^{+0.00098}_{-0.00095}$ &      $3360.0516^{+0.0018}_{-0.0018}$ &       $3.65^{+0.25}_{-0.75}$ &      $4.24^{+0.64}_{-0.84}$ &  0.77 &      5232 &      $3.56^{+0.54}_{-0.70}$ &       27.2                 &    K2-198b \citep{mayo:2018} \\                                                                        
 212768333.02 &  16.825 &         $7.44957^{+0.00067}_{-0.00068}$ &      $3349.0808^{+0.0034}_{-0.0034}$ &       $2.86^{+0.56}_{-0.22}$ &      $2.80^{+0.29}_{-0.30}$ &  0.77 &      5232 &      $2.34^{+0.24}_{-0.25}$ &       81.9                 &     Candidate from \cite{pope:2016} \\                                                                       
 212771557.01 &  13.950 &            $8.4902^{+0.0014}_{-0.0014}$ &      $3349.4717^{+0.0047}_{-0.0048}$ &       $2.55^{+0.32}_{-0.21}$ &      $2.56^{+0.26}_{-0.23}$ &  0.86 &      5530 &      $2.39^{+0.25}_{-0.22}$ &           99               &    ---   \\                                                                                                
 212779563.01 &   9.945 &         $6.00123^{+0.00012}_{-0.00018}$ &   $3352.36041^{+0.00101}_{-0.00079}$ &    $1.272^{+0.102}_{-0.031}$ &      $2.73^{+0.11}_{-0.12}$ &  0.69 &      4688 &   $2.064^{+0.088}_{-0.097}$ &       62.5                 &      Wolf 503 (Peterson et al., in prep.) \\                                                                                            
 212779596.01 &  11.930 &         $7.37416^{+0.00023}_{-0.00023}$ &      $3348.6147^{+0.0011}_{-0.0011}$ &    $2.361^{+0.128}_{-0.091}$ &      $4.02^{+0.25}_{-0.19}$ &  0.67 &      4772 &      $2.93^{+0.18}_{-0.14}$ &       48.2                 &     K2-199b \citep{mayo:2018}\\                                                                        
 212779596.02 &  11.930 &         $3.22575^{+0.00014}_{-0.00014}$ &      $3346.9032^{+0.0017}_{-0.0017}$ &    $1.872^{+0.151}_{-0.090}$ &      $2.58^{+0.16}_{-0.14}$ &  0.67 &      4772 &      $1.88^{+0.12}_{-0.10}$ &          145               &     K2-199c \citep{mayo:2018} \\                                                                          
 212803289.01 &  11.014 &        $18.24605^{+0.00083}_{-0.00090}$ &      $3349.7141^{+0.0016}_{-0.0016}$ &   $10.905^{+0.085}_{-0.076}$ &   $3.738^{+0.075}_{-0.047}$ &  2.59 &      6560 &     $10.57^{+0.38}_{-0.35}$ &          422               &     K2-99b \citep{smith:2017} \\                                                                          
 212813907.01 &  14.070 &         $6.72526^{+0.00031}_{-0.00033}$ &      $3350.5430^{+0.0016}_{-0.0016}$ &       $0.82^{+0.17}_{-0.13}$ &      $5.56^{+1.27}_{-0.59}$ &  0.79 &      5007 &      $4.79^{+1.10}_{-0.52}$ &       82.1                 &     \nd \\                                                                                             
 212870185.01 &  13.149 &         $6.11665^{+0.00044}_{-0.00044}$ &      $3347.9964^{+0.0027}_{-0.0026}$ &       $2.54^{+0.28}_{-0.18}$ &      $3.04^{+0.28}_{-0.25}$ &  1.12 &      5587 &      $3.73^{+0.36}_{-0.32}$ &          258               &     ---  \\                                                                                               
 251554286.01 &  12.091 &        $15.46659^{+0.00066}_{-0.00064}$ &      $3356.8506^{+0.0011}_{-0.0012}$ &       $3.55^{+0.37}_{-0.42}$ &      $4.44^{+0.50}_{-0.80}$ &  0.98 &      5657 &      $4.73^{+0.56}_{-0.85}$ &       60.0                 &     ---  \\                                                                                            
 251582120.01 &  15.175 &      $0.509967^{+0.000055}_{-0.000051}$ &      $3346.9256^{+0.0029}_{-0.0043}$ &       $3.25^{+0.56}_{-0.59}$ &      $4.72^{+0.81}_{-0.44}$ &  1.25 &      5997 &      $6.49^{+1.18}_{-0.78}$ &    10946                   &     \nd \\                                                                                          
 251590700.01 &  13.302 &         $5.82105^{+0.00097}_{-0.00100}$ &      $3347.5528^{+0.0058}_{-0.0058}$ &          $6.1^{+3.3}_{-6.1}$ &      $6.40^{+0.78}_{-0.50}$ &  0.86 &      5247 &         $6.1^{+3.9}_{-3.8}$ &        138                 &    Low $\rho_{*,circ}$.   \\                                                                                          
 \enddata
\end{deluxetable*}

\clearpage
\end{turnpage}

\clearpage
\begin{turnpage}
\begin{deluxetable}{lr|lrllrrl|llrl|lll}
  \tabletypesize{\scriptsize}\tablecaption{Stellar Parameters}\tablehead{
\multicolumn{2}{c}{} & \multicolumn{7}{|c|}{TRES} & \multicolumn{4}{c|}{HIRES\tablenotemark{a}} & \multicolumn{3}{c}{SpeX\tablenotemark{b}} \\
EPIC &    $Kp$ &         BJD$_{UTC}$\tablenotemark{c} &   S/N\tablenotemark{d} & \teff & \logg &  [M/H] & \vsini\tablenotemark{e} &         RV\tablenotemark{f} &  \teff & \logg & [Fe/H] &    \vsini &   SpT & \teff & \logg \\
& [mag] & [days] & & [K] & [dex] & [dex] & [km~s$^{-1}$] & km~s$^{-1}$ & [K] & [dex] & [dex] & [km~s$^{-1}$] & & [K]  & [dex] \\
}\startdata
  212428509 &  12.5 &             \nd &   \nd &       \nd &       \nd &      \nd &        \nd &          \nd &   5697 &    4.25 &$-0.42$ &  1.7 &       \nd   &  \nd   &    \nd  \\
  212435047 &  12.4 &             \nd &   \nd &       \nd &       \nd &      \nd &        \nd &          \nd &   5750 &    4.29 &   0.01 &  2.0 &       \nd   &  \nd   &    \nd  \\
  212460519 &  12.4 &             \nd &   \nd &       \nd &       \nd &      \nd &        \nd &          \nd &   4226\tablenotemark{g} &     \nd &$-0.17$ &  \nd &       \nd   &  \nd   &    \nd  \\
  212496592 &   13.0 &  2457435.973127 &  25.4 &      5177 &      4.57 &   $0.31$ &        2.8 &   $-9.060$ &      \nd &     \nd &    \nd &  \nd &       \nd   &  \nd   &    \nd  \\
  212521166 &  11.6 &  2457436.932008 &  27.7 &      4912 &      4.57 &  $-0.29$ &        1.7 &  $-21.573$ &   4895   &    4.64 &$-0.24$ &  1.9 &        K2V  &  4841  & 4.63  \\
  212554013 &  14.7  &  \nd            &  \nd  &       \nd &       \nd &      \nd &        \nd &        \nd &      \nd &     \nd &    \nd &  \nd &       K3V   &  4388  & 4.64 \\
  212572439 &   12.8 &  2457442.944484 &  16.4 &      5123 &      4.57 &   $0.45$ &        6.3 &   $13.835$ &      \nd &     \nd &    \nd &  \nd &       K2V   &  4972  & 4.59 \\
  212580872 &   13.0 &  2457493.742254 &  30.5 &      5612 &      4.45 &   $0.20$ &        3.5 &  $-16.946$ &      \nd &     \nd &    \nd &  \nd &       \nd   &  \nd   &    \nd  \\
  212586030 &  11.7 &             \nd &   \nd &       \nd &       \nd &      \nd &        \nd &          \nd & 4865   &    3.37 &   0.38 &  3.5 &       \nd   &  \nd   &    \nd  \\
  212587672 &  12.2 &             \nd &   \nd &       \nd &       \nd &      \nd &        \nd &          \nd &   5948 &    4.49 &$-0.21$ &  2.1 &       \nd   &  \nd   &    \nd  \\
  212619190 &  12.8 &  2458273.731631 &   28.7 &       5648 &       4.33 &      0.04 &    4.6 &     29.555   &   \nd &    \nd & \nd &  \nd &       \nd   &  \nd   &    \nd  \\
  212628254 &   9.7 &  2458261.733258 &  51.6 &      5833 &      4.40 &  $-0.01$ &        3.0 &  $-28.074$ &     5827\tablenotemark{h} &    4.31\tablenotemark{h} &   0.04\tablenotemark{h} &  \nd &       \nd   &  \nd   &    \nd  \\
  212628477\tablenotemark{i} 
              & 12.5 &  2458274.706803 & 27.5  & \nd              & \nd           &  \nd       & \nd        & \nd &        \nd &        \nd &      \nd &     \nd &    \nd &  \nd &       \nd   \\
  212634172 &  14.8  &  \nd            &  \nd  &       \nd &       \nd &      \nd &        \nd &        \nd &      \nd &     \nd &    \nd &  \nd &       M3V   &  3412  & 4.86 \\  
  212651213\tablenotemark{i}
            &   10.8 &  2457439.912117 &  52.2 &       \nd &       \nd &      \nd &        \nd &        \nd &      \nd &     \nd &    \nd &  \nd &       \nd   &  \nd   &    \nd  \\ 
      ''    &  ''   &  2457448.969440 &  41.0 &       \nd &       \nd &      \nd &        \nd &        \nd &      \nd &     \nd &    \nd &  \nd &       \nd   &  \nd   &    \nd  \\
      ''    &  ''   &  2457449.945082 &  38.5 &       \nd &       \nd &      \nd &        \nd &        \nd &      \nd &     \nd &    \nd &  \nd &       \nd   &  \nd   &    \nd  \\
      ''    &  ''   &  2457450.917452 &  37.7 &       \nd &       \nd &      \nd &        \nd &        \nd &      \nd &     \nd &    \nd &  \nd &       \nd   &  \nd   &    \nd  \\
      ''    &  ''   &  2457451.909447 &  37.3 &       \nd &       \nd &      \nd &        \nd &        \nd &      \nd &     \nd &    \nd &  \nd &       \nd   &  \nd   &    \nd  \\
      ''    &   ''  &  2457452.902042 &  25.8 &       \nd &       \nd &      \nd &        \nd &        \nd &      \nd &     \nd &    \nd &  \nd &       \nd   &  \nd   &    \nd  \\
      ''    &   ''  &  2457454.892102 &  36.6 &       \nd &       \nd &      \nd &        \nd &        \nd &      \nd &     \nd &    \nd &  \nd &       \nd   &  \nd   &    \nd  \\
      ''    &   ''  &  2457470.863085 &  37.6 &       \nd &       \nd &      \nd &        \nd &        \nd &      \nd &     \nd &    \nd &  \nd &       \nd   &  \nd   &    \nd  \\
  212651234\tablenotemark{g}
            &   11.1 &  2457439.929578 &  49.3 &      4902 &      3.50 &   $0.23$ &        2.6 &  $-15.508$ &      \nd &     \nd &    \nd &  \nd &       \nd   &  \nd   &    \nd  \\
      ''    &  ''   &  2457448.983742 &  27.1 &      4853 &      3.34 &   $0.24$ &        2.9 &  $-15.376$ &      \nd &     \nd &    \nd &  \nd &       \nd   &  \nd   &    \nd  \\
      ''    &  ''   &  2457452.911059 &  15.1 &      4901 &      3.46 &   $0.39$ &        4.9 &  $-15.350$ &      \nd &     \nd &    \nd &  \nd &       \nd   &  \nd   &    \nd  \\
      ''    &  ''   &  2457466.925434 &  32.5 &      5078 &      3.94 &   $0.35$ &        2.0 &  $-15.399$ &      \nd &     \nd &    \nd &  \nd &       \nd   &  \nd   &    \nd  \\
      ''    &  ''   &  2457504.855779 &  23.4 &      4807 &      3.22 &   $0.26$ &        3.9 &  $-15.421$ &      \nd &     \nd &    \nd &  \nd &       \nd   &  \nd   &    \nd  \\
      ''    &  ''   &  2457511.879130 &  20.4 &      4861 &      3.42 &   $0.30$ &        3.9 &  $-15.631$ &      \nd &     \nd &    \nd &  \nd &       \nd   &  \nd   &    \nd  \\
  212686205 &   12.3 &  2457435.907480 &  28.2 &      4635 &      4.70 &  $-0.23$ &        2.3 &  $-12.053$ &      \nd &     \nd &    \nd &  \nd &       K4V   &  4470  & 4.51  \\
  212689874 &  12.3 &  2457434.882603 &  29.2 &      5714 &      4.55 &  $-0.09$ &        3.0 &  $-14.721$ &   5644   &    4.36 &$-0.12$ &  1.7 &       \nd   &  \nd   &    \nd  \\
  212697709 &  12.2 &  2457439.975173 &  40.1 &      5785 &      4.45 &   $0.31$ &        3.1 &  $-21.995$ &   5719   &    4.28 &   0.28 &  1.6 &       \nd   &  \nd   &    \nd  \\
        ''  &   ''  &  2457439.997975 &  39.6 &      5733 &      4.38 &   $0.31$ &        3.6 &  $-22.019$ &      \nd &     \nd &    \nd &  \nd &       \nd   &  \nd   &    \nd  \\
        ''  &   ''  &  2457475.857401 &  34.1 &      5796 &      4.46 &   $0.32$ &        3.4 &  $-21.918$ &      \nd &     \nd &    \nd &  \nd &       \nd   &  \nd   &    \nd  \\
  212705192\tablenotemark{i}
            &   11.7 &  2457439.893014 &  53.7 &       \nd &       \nd &      \nd &       \nd &        \nd &      \nd &     \nd &    \nd &  \nd &       \nd   &  \nd   &    \nd  \\
  212735333 &  12.0 &  2457439.870513 &  44.5 &      5671 &      4.57 &  $-0.01$ &       2.3  &   $-6.591$ &   5660   &    4.50 &   0.09 &  1.3 &       \nd   &  \nd   &    \nd  \\
  212768333 &   11.0 &  2457439.037432 &  54.1 &      5247 &      4.61 &  $-0.16$ &       5.2 &    $2.071$ &      \nd &     \nd &    \nd &  \nd &       \nd   &  \nd   &    \nd  \\
  212779596 &  11.9 &  2457437.046415 &  25.6 &      4652 &      4.63 &  $-0.21$ &        2.1 &    $0.092$ &   4507\tablenotemark{g}   &     \nd &$-0.04$ &  \nd &        K5V  &  4731  & 4.62 \\
  212782836 &  11.6 &             \nd &   \nd &       \nd &       \nd &      \nd &        \nd &          \nd & 5418   &    4.48 &$-0.42$ &  1.1 &       \nd   &  \nd   &    \nd  \\
  212779563 &   9.8 &  2458261.725801 &  45.3 &      4640 &      4.68 &  $-0.47$ &        0.8 &  $-46.629$ &     4568\tablenotemark{g,h} &     \nd &    \nd &  \nd &       \nd   &  \nd   &    \nd  \\
  212803289 &  11.0 &  2457437.035094 &  42.0 &      6048 &      3.79 &   $0.11$ &       11.1 &   $-2.778$ &   6102   &    3.96 &   0.20 &  10.0 &      \nd   &  \nd   &    \nd  \\
        ''  &   ''  &  2457447.858765 &  37.7 &      5906 &      3.58 &   $0.03$ &       11.5 &   $-2.559$ &      \nd &     \nd &    \nd &  \nd &       \nd   &  \nd   &    \nd  \\
        ''  &   ''  &  2457475.842684 &  29.0 &      6105 &      3.87 &   $0.30$ &       12.0 &   $-2.554$ &      \nd &     \nd &    \nd &  \nd &       \nd   &  \nd   &    \nd  \\
  251539584\tablenotemark{i}  &
            10.8 & 2458274.726575        & 29.1 &       \nd  &       \nd  &  \nd    & \nd               & \nd &      \nd &     \nd &    \nd &  \nd &       \nd   &  \nd   &    \nd  \\
  '' & ''            & 2458276.738180 & 31.3 & \nd  & \nd  &  \nd    & \nd & \nd &      \nd &     \nd &    \nd &  \nd &       \nd   &  \nd   &    \nd   \\
  251539609\tablenotemark{i} &
       11.0    & 2458275.698478 & 35.3 & \nd  & \nd  &  \nd    & \nd & \nd   \nd &     \nd &    \nd &  \nd &       \nd   &  \nd   &    \nd  \\
 '' & ''             & 2458276.730773 & 30.1 & \nd  & \nd  &  \nd    & \nd & \nd   \nd &     \nd &    \nd &  \nd &       \nd   &  \nd   &    \nd  \\
 251554286 &  12.1 &  2458275.686467 &  30.5 &      5548 &      4.44 & $-0.10$ &         1.0 &   $4.560$  &      \nd &     \nd &    \nd &  \nd &       \nd   &  \nd   &    \nd

        \enddata\label{tab:spec}

  \tablenotetext{a}{
    HIRES data and analysis described by \citep{petigura:2018}.
  }
  \tablenotetext{b}{
    SpeX data and analysis described by \citep{dressing:2017a}.
  }
  \tablenotetext{c}{
    Date of TRES observation. 
  }
  \tablenotetext{d}{
    Signal-to-noise ratio per resolution element in the wavelength
    range $5060$\ to $5315$\,\AA.
  }
  \tablenotetext{e}{
    SPC measures the broadening from an edge-on rotator with a fixed
    macroturbulent velocity of $1\,\kms$. Different values of
    macroturbulence may bias this value for slow rotators. As such, we
    caution against interpreting this value as \vsini\ without further
    analysis.
  }
  \tablenotetext{f}{
    The RVs reported here have been shifted onto the IAU scale using
    standard star velocities, on which, e.g., HD\,182488 has an
    absolute RV of $-21.508$ \citep{nidever:2002}. The uncertainties
    of the reconnaissance RVs on the TRES native system are typically
    on the order of $50$\,\ms (also affected by \teff, S/N
    and \vsini), though the offset to the absolute scale carries
    similar uncertainty.
  }
  \tablenotetext{g}{
        Star too cool for SpecMatch analysis \citep[see ][]{petigura:2018}.
  }
  \tablenotetext{h}{
        Star observed with APF instead of HIRES, but stellar parameters inferred using the same approach as described in \cite{petigura:2018}.
  }
  
  \tablenotetext{i}{Multi-lined spectrum. 
  }
\end{deluxetable}


\clearpage
\end{turnpage}

\clearpage
\LongTables 
\begin{deluxetable}{rrlrrrl}
\tabletypesize{\scriptsize}\tablecaption{Eclipsing Binaries}\tablehead{
     &    $Kp$ &             Epoch &     $P$ &  $T_{14}$ &  $(R_P/R_*)^2$ &                                                     \\
EPIC  & [mag]  & [BJD$_{TDB}$]  & [d] & [d] &  & comments \\
}\startdata
 212628098 &  13.259 &  2458180.89299 &  4.352574 &  0.067307 &  0.042013 &  \nd \\
 212651213 &  10.796 &  2458180.35821 &  2.538338 &  0.144896 &  0.044374 &  V-shaped, large radius \\
 212658818 &  12.070 &  2458180.48591 &  2.321117 &  0.066364 &  0.000868 &  blend because transit depth not consistent (not on target) \\
 212757601 &  16.825 &  2458179.98367 &  1.017967 &  0.057751 &  0.012362 &  Jovian planet around small star? 7.7 $R_\oplus$ \\
 212769367 &  17.911 &  2458199.34193 &  20.225392 &  0.258937 &  0.021858 &  \nd \\
 212769682 &  18.382 &  2458199.34810 &  20.230002 &  0.276014 &  0.041586 &  GAIA parallax $<$1 mas \\
 212871068 &  18.318 &  2458182.72856 &  8.744013 &  0.183117 &  0.140517 &  \\ 
 212884586 &  17.700 &  2458180.15931 &  2.882978 &  0.049651 &  0.011687 &  \nd \\
 251810686 &  10.865 &  2458180.36230 &  2.537920 &  0.164611 &  0.059434 &  bad aperture; \cite{rappaport:2016} \\
 212581374 &  10.292 &  2458180.14795 &  0.784498 &  0.157174 &  0.003875 &  \nd \\
 212406350 &  13.923 &  2458179.72331 &  0.833679 &  0.083508 &  0.096367 &  \nd \\
 212409856 &  13.446 &  2458179.83675 &  0.531704 &  0.078146 &  0.159770 &  \nd \\
 212417656 &  12.745 &  2458179.74444 &  0.815627 &  0.136918 &  0.023504 &  \nd \\
 212420474 &  13.442 &  2458179.83016 &  0.600579 &  0.066488 &  0.044711 &  \nd \\
 212420510 &  14.632 &  2458179.82589 &  0.600656 &  0.077941 &  0.145720 &  contact \\
 212421319 &  16.407 &  2458182.18746 &  5.528665 &  0.239914 &  0.014466 &  odd-even, wrong period \\
 212421673 &  13.172 &  2458187.99492 &  28.248155 &  0.446599 &  0.003888 &  \nd \\
 212426112 &  13.150 &  2458179.89122 &  1.530195 &  0.072284 &  0.035180 &  \nd \\
 212428509 &  12.483 &  2458180.30248 &  2.667940 &  0.080248 &  0.007745 &  odd-even effect \\
 212435964 &  14.080 &  2458193.11111 &  25.184817 &  0.201155 &  0.234665 &  \nd \\
 212439709 &  14.352 &  2458180.15803 &  1.218136 &  0.066728 &  0.056980 &  contact, same as 1 \\
 212442107 &  15.821 &  2458180.02735 &  0.546059 &  0.074620 &  0.273964 &  \nd \\
 212442408 &  11.778 &  2458180.41810 &  0.909676 &  0.123028 &  0.255280 &  \nd \\
 212453473 &  13.957 &  2458181.97486 &  2.756129 &  0.150371 &  0.323040 &  \nd \\
 212454161 &  15.225 &  2458180.76138 &  22.334245 &  0.610513 &  0.022610 &  \nd \\
 212455982 &  14.140 &  2458180.67276 &  1.620017 &  0.242113 &  0.107147 &  \nd \\
 212456583 &  13.429 &  2458182.17512 &  2.877393 &  0.164731 &  0.161885 &  \nd \\
 212460623 &  9.086 &  2458179.98967 &  0.492488 &  0.086255 &  0.000156 &  \nd \\
 212465919 &  15.159 &  2458180.05317 &  0.569619 &  0.081742 &  0.230555 &  contacting \\
 212468149 &  14.814 &  2458179.86667 &  0.688366 &  0.059358 &  0.114282 &  \nd \\
 212473154 &  8.980 &  2458181.23537 &  1.816975 &  0.083992 &  0.002040 &  \nd \\
 212481328 &  13.090 &  2458179.55397 &  3.417361 &  0.105410 &  0.048337 &  \nd \\
 212488008 &  10.633 &  2458189.49044 &  11.334688 &  0.070855 &  0.001533 &  \nd \\
 212491978 &  14.025 &  2458179.95415 &  0.535811 &  0.062105 &  0.071267 &  contact ,same as 1 \\
 212497267 &  12.282 &  2458182.01007 &  3.744355 &  0.180382 &  0.285638 &  \nd \\
 212499716 &  13.748 &  2458180.06238 &  0.874745 &  0.035389 &  0.001790 &  \nd \\
 212502064 &  9.671 &  2458179.70262 &  0.560679 &  0.088106 &  0.049133 &  contact \\
 212504385 &  13.842 &  2458179.91896 &  0.826894 &  0.122608 &  0.249751 &  \nd \\
 212509737 &  11.997 &  2458179.59591 &  2.343356 &  0.059597 &  0.008323 &  \nd \\
 212511920 &  13.209 &  2458179.99753 &  0.572508 &  0.076707 &  0.097044 &  contact \\
 212512022 &  16.643 &  2458179.89864 &  0.514313 &  0.124243 &  0.002423 &  contact \\
 212518838 &  15.643 &  2458179.80762 &  0.651904 &  0.081742 &  0.198824 &  contact \\
 212523277 &  17.547 &  2458179.75820 &  13.538932 &  0.114329 &  0.087378 &  \nd \\
 212527975 &  13.708 &  2458179.68204 &  0.517780 &  0.081742 &  0.157632 &  contact \\
 212530520 &  15.411 &  2458180.29465 &  0.808487 &  0.093941 &  0.118684 &  contact \\
 212535959 &  13.803 &  2458190.36673 &  17.733194 &  0.292331 &  0.111249 &  \nd \\
 212537106 &  12.982 &  2458181.36656 &  9.263450 &  0.273879 &  0.163254 &  \nd \\
 212540174 &  14.869 &  2458179.57468 &  0.527054 &  0.040555 &  0.056895 &  contact \\
 212540985 &  13.574 &  2458179.85092 &  0.548227 &  0.078714 &  0.035505 &  \nd \\
 212541386 &  14.231 &  2458181.74987 &  3.630331 &  0.091115 &  0.074444 &  \nd \\
 212545451 &  15.672 &  2458179.79113 &  1.133767 &  0.154570 &  0.450641 &  \nd \\
 212545602 &  16.209 &  2458180.61219 &  1.756713 &  0.220238 &  0.670509 &  \nd \\
 212546446 &  14.369 &  2458179.68614 &  0.655294 &  0.081742 &  0.133002 &  contact \\
 212553193 &  15.314 &  2458179.68060 &  0.570422 &  0.079264 &  0.233006 &  \nd \\
 212559866 &  11.864 &  2458184.00383 &  19.702223 &  0.383548 &  0.248986 &  \nd \\
 212560752 &  12.839 &  2458179.91313 &  0.582783 &  0.081742 &  0.097117 &  \nd \\
 212566769 &  13.331 &  2458189.13230 &  14.301229 &  0.323096 &  0.039127 &  \nd \\
 212567829 &  18.076 &  2458180.10226 &  0.841796 &  0.119074 &  0.284914 &  \nd \\
 212570257 &  12.523 &  2458179.69542 &  0.610230 &  0.055085 &  0.070548 &  secondary of contacting \\
 212577519 &  14.234 &  2458180.54062 &  0.980712 &  0.077982 &  0.115798 &  contact \\
 212579164 &  13.632 &  2458182.64844 &  18.155715 &  0.137503 &  0.230781 &  46 $R_\oplus$ \\
 212580081 &  18.233 &  2458180.41422 &  1.491851 &  0.088955 &  0.692969 &  35 $R_\oplus$ \\
 212580230 &  12.838 &  2458179.96998 &  0.563909 &  0.081742 &  0.367660 &  Contact \\
 212586717 &  13.875 &  2458181.71797 &  4.295939 &  0.087219 &  0.012705 &  \nd \\
 212601505 &  14.486 &  2458179.96618 &  0.724453 &  0.035719 &  0.020973 &  \nd \\
 212609851 &  15.164 &  2458179.82750 &  0.642765 &  0.057191 &  0.223025 &  \nd \\
 212611243 &  14.163 &  2458179.94634 &  0.726623 &  0.077036 &  0.097420 &  \nd \\
 212612033 &  18.300 &  2458179.98494 &  1.049595 &  0.091376 &  0.022397 &  \nd \\
 212613128 &  13.861 &  2458180.19045 &  0.759210 &  0.070657 &  0.213789 &  \nd \\
 212615099 &  15.660 &  2458192.20124 &  16.397313 &  0.105083 &  0.122559 &  \nd \\
 212617879 &  12.316 &  2458179.84646 &  2.210766 &  0.153759 &  0.142075 &  \nd \\
 212627712 &  13.265 &  2458186.21980 &  19.913432 &  0.145782 &  0.165860 &  107 $R_\oplus$ \\
 212629807 &  15.143 &  2458179.90970 &  0.501935 &  0.081742 &  0.206343 &  contact \\
 212631911 &  15.546 &  2458179.98736 &  0.520852 &  0.078445 &  0.333555 &  \nd \\
 212634594 &  15.202 &  2458184.28069 &  6.401944 &  0.145015 &  0.212873 &  \nd \\
 212641218 &  14.993 &  2458179.98311 &  1.049606 &  0.076901 &  0.001691 &  \nd \\
 212644753 &  9.422 &  2458179.97694 &  1.049846 &  0.097062 &  0.041131 &  \nd \\
 212651213 &  10.796 &  2458191.53766 &  13.196894 &  0.199239 &  0.010896 &  \cite{rappaport:2016}\\
 212651234 &  11.139 &  2458180.35324 &  2.538731 &  0.123252 &  0.008702 &  \cite{rappaport:2016}; 30.5 $R_\oplus$ \\
 212652663 &  14.819 &  2458180.77106 &  1.669747 &  0.102005 &  0.228074 &  \nd \\
 212654750 &  13.917 &  2458179.88743 &  0.529294 &  0.081742 &  0.413695 &  contact \\
 212657659 &  17.470 &  2458180.01607 &  0.546679 &  0.055120 &  0.014074 &  contact \\
 212666524 &  14.293 &  2458179.90638 &  0.670516 &  0.081742 &  0.121268 &  \nd \\
 212666639 &  15.366 &  2458179.54065 &  0.541019 &  0.079310 &  0.301795 &  contact \\
 212667298 &  12.902 &  2458179.54657 &  0.606965 &  0.081742 &  0.435121 &  contact \\
 212671857 &  13.697 &  2458180.24217 &  0.727391 &  0.068894 &  0.139981 &  \nd \\
 212679798 &  14.846 &  2458180.12895 &  1.834750 &  0.073377 &  0.033351 &  \nd \\
 212686943 &  13.774 &  2458181.02088 &  1.578709 &  0.165925 &  0.064449 &  \nd \\
 212687040 &  13.475 &  2458180.27371 &  1.852983 &  0.106111 &  0.205153 &  \nd \\
 212689699 &  17.593 &  2458180.07219 &  0.518523 &  0.130845 &  0.013282 &  contact \\
 212690087 &  14.746 &  2458180.09903 &  0.786832 &  0.114912 &  0.042193 &  \nd \\
 212691727 &  12.657 &  2458184.17922 &  12.862016 &  0.201678 &  0.050839 &  \nd \\
 212695400 &  15.403 &  2458180.22806 &  0.848459 &  0.065686 &  0.215148 &  \nd \\
 212697951 &  12.582 &  2458180.27911 &  1.912398 &  0.114449 &  0.259949 &  star spot causes modulation \\
 212701118 &  12.691 &  2458179.72465 &  2.434027 &  0.144225 &  0.661748 &  \nd \\
 212702889 &  14.558 &  2458179.93264 &  0.631071 &  0.056983 &  0.052287 &  \nd \\
 212705192 &  11.728 &  2458181.41157 &  2.268360 &  0.048411 &  0.005948 &  odd-even effect, double-lined \\
 212705508 &  14.415 &  2458180.05063 &  0.603816 &  0.044304 &  0.003131 &  \nd \\
 212707624 &  13.179 &  2458182.00981 &  3.604588 &  0.207304 &  0.106715 &  \nd \\
 212708296 &  15.906 &  2458180.26857 &  0.803247 &  0.100811 &  0.466097 &  \nd \\
 212708783 &  10.386 &  2458179.95230 &  2.253755 &  0.142294 &  0.118586 &  \nd \\
 212710571 &  17.458 &  2458179.95368 &  2.253558 &  0.104992 &  0.012538 &  \nd \\
 212712870 &  15.304 &  2458179.96661 &  0.494226 &  0.069594 &  0.249001 &  \nd \\
 212716448 &  18.478 &  2458180.01069 &  0.546752 &  0.058736 &  0.062706 &  same as 1 \\
 212723069 &  14.817 &  2458186.05758 &  11.495130 &  0.232389 &  0.037574 &  \nd \\
 212723581 &  15.961 &  2458180.00972 &  0.600845 &  0.066764 &  0.124436 &  same signal as 1 \\
 212733831 &  14.786 &  2458179.70777 &  0.732994 &  0.081742 &  0.117807 &  \nd \\
 212734205 &  17.588 &  2458181.12287 &  4.965604 &  0.493681 &  0.397380 &  \nd \\
 212737890 &  15.875 &  2458179.84702 &  0.880552 &  0.105444 &  0.127097 &  \nd \\
 212740148 &  13.996 &  2458180.15919 &  0.741042 &  0.030996 &  0.011375 &  \nd \\
 212741343 &  15.933 &  2458180.05956 &  0.580501 &  0.054682 &  0.100483 &  contact \\
 212746282 &  12.518 &  2458179.85030 &  0.595119 &  0.081742 &  0.093743 &  contact \\
 212747879 &  15.717 &  2458179.97540 &  0.705760 &  0.081742 &  0.331363 &  \nd \\
 212748031 &  15.678 &  2458180.36357 &  0.887395 &  0.037098 &  0.005056 &  \nd \\
 212751079 &  13.700 &  2458179.62410 &  0.595131 &  0.142401 &  0.264229 &  \nd \\
 212751916 &  13.890 &  2458180.64439 &  15.715606 &  0.097758 &  0.004367 &  \nd \\
 212759326 &  13.892 &  2458182.52706 &  3.376283 &  0.117698 &  0.076310 &  \nd \\
 212770429 &  11.153 &  2458199.35119 &  20.225506 &  0.342386 &  0.210533 &  75 $R_\oplus$ \\
 212771092 &  17.554 &  2458180.04000 &  0.613816 &  0.081742 &  0.513770 &  \nd \\
 212771522 &  14.105 &  2458180.36577 &  0.964855 &  0.036899 &  0.002141 &  \nd \\
 212773272 &  14.965 &  2458182.45629 &  4.681890 &  0.080497 &  0.043560 &  \nd \\
 212773309 &  11.391 &  2458182.45642 &  4.681764 &  0.093543 &  0.074791 &  \nd \\
 212781530 &  15.601 &  2458180.03084 &  0.574416 &  0.081742 &  0.518721 &  contact \\
 212781903 &  13.952 &  2458179.93093 &  0.516312 &  0.081742 &  0.057071 &  \nd \\
 212786474 &  14.472 &  2458179.57656 &  9.271273 &  0.151254 &  0.429256 &  \nd \\
 212789681 &  13.740 &  2458179.55289 &  0.497467 &  0.116872 &  0.000516 &  contact \\
 212796590 &  16.506 &  2458179.97098 &  0.555792 &  0.144363 &  0.009497 &  contact \\
 212801119 &  12.771 &  2458180.11071 &  0.591442 &  0.045596 &  0.019034 &  \nd \\
 212801667 &  11.911 &  2458186.41163 &  23.274142 &  0.214440 &  0.075892 &  \nd \\
 212805198 &  14.422 &  2458180.96489 &  3.228788 &  0.086784 &  0.079089 &  \nd \\
 212812349 &  13.712 &  2458185.62953 &  8.167374 &  0.174965 &  0.069996 &  \nd \\
 212814517 &  15.896 &  2458179.76158 &  0.624914 &  0.079529 &  0.314121 &  \nd \\
 212822491 &  11.078 &  2458186.08017 &  14.321271 &  0.265478 &  0.171877 &  \nd \\
 212824416 &  16.638 &  2458179.85284 &  0.590807 &  0.057018 &  0.134113 &  contact EB; secondary \\
 212826509 &  16.297 &  2458180.41915 &  0.988762 &  0.113296 &  0.311666 &  \nd \\
 212827749 &  13.358 &  2458185.76643 &  11.345548 &  0.187133 &  0.207902 &  \nd \\
 212828964 &  16.170 &  2458179.90943 &  0.646399 &  0.142256 &  0.001916 &  contact \\
 212834326 &  15.554 &  2458180.10438 &  0.780977 &  0.079370 &  0.242254 &  \nd \\
 212837770 &  16.663 &  2458180.22595 &  0.850575 &  0.064098 &  0.263615 &  \nd \\
 212839815 &  12.874 &  2458180.59961 &  4.441165 &  0.198630 &  0.037661 &  \nd \\
 212842049 &  16.894 &  2458181.48623 &  3.289052 &  0.066265 &  0.062749 &  \nd \\
 212842366 &  12.081 &  2458179.58419 &  0.543994 &  0.059710 &  0.018823 &  \nd \\
 212854191 &  12.566 &  2458180.39309 &  0.868807 &  0.099834 &  0.046954 &  contact \\
 212864075 &  11.826 &  2458180.11467 &  0.729410 &  0.071462 &  0.015258 &  \nd \\
 212866286 &  12.702 &  2458180.51003 &  4.717350 &  0.245227 &  0.178060 &  \nd \\
 212869892 &  12.392 &  2458179.99254 &  0.814852 &  0.057258 &  0.008050 &  \nd \\
 212872008 &  14.464 &  2458180.76477 &  1.311925 &  0.107024 &  0.102602 &  \nd \\
 212872519 &  18.895 &  2458180.02866 &  1.361929 &  0.188677 &  0.316683 &  \nd \\
 212878430 &  18.479 &  2458179.64683 &  0.511345 &  0.081742 &  0.086995 &  contact \\
 212884295 &  16.098 &  2458180.05753 &  0.632894 &  0.082281 &  0.151918 &  contact \\
 212885442 &  15.582 &  2458179.58563 &  0.626888 &  0.081742 &  0.192118 &  \nd \\
 251505087 &  16.021 &  2458180.01374 &  0.744603 &  0.080170 &  0.204046 &  \nd \\
 251505480 &  18.300 &  2458179.54528 &  0.622504 &  0.080448 &  0.117676 &  contact \\
 251505499 &  9.619 &  2458179.54539 &  0.622507 &  0.081742 &  0.278995 &  contact \\
 251508456 &  15.216 &  2458179.90526 &  0.774116 &  0.142628 &  0.773576 &  \nd \\
 251508975 &  16.979 &  2458179.93148 &  0.583320 &  0.081742 &  0.142980 &  \nd \\
 251512942 &  14.262 &  2458179.54192 &  0.546855 &  0.081742 &  0.249001 &  contacting \\
 251523672 &  16.201 &  2458179.84407 &  0.594784 &  0.043602 &  0.153440 &  contact \\
 251524025 &  16.805 &  2458179.79873 &  0.638134 &  0.073617 &  0.386702 &  \nd \\
 251539042 &  15.597 &  2458179.53378 &  0.561767 &  0.076747 &  0.249001 &  \nd \\
 251543556 &  13.596 &  2458179.96760 &  0.498006 &  0.049089 &  0.018157 &  \nd \\
 251551459 &  16.526 &  2458179.76260 &  0.938771 &  0.083508 &  0.235088 &  \nd \\
 251566115 &  12.519 &  2458182.48929 &  11.850868 &  0.127530 &  0.072908 &  \nd \\
 251567015 &  16.442 &  2458179.68328 &  0.558434 &  0.073032 &  0.111879 &  contact \\
 251571270 &  17.339 &  2458179.61675 &  0.645707 &  0.048994 &  0.425897 &  \nd \\
 251575183 &  18.642 &  2458179.89846 &  0.515838 &  0.070330 &  0.116968 &  \nd \\
 251600179 &  17.983 &  2458179.74495 &  0.668258 &  0.055939 &  0.071262 &  \nd \\
 251606815 &  15.059 &  2458179.53572 &  0.514761 &  0.081742 &  0.405411 &  \nd \\
 251612064 &  15.053 &  2458179.72566 &  0.519174 &  0.081742 &  0.367738 &  \nd \\
 251613109 &  17.532 &  2458180.09242 &  0.603096 &  0.075259 &  0.282421 &  \nd \\
 251628925 &  12.632 &  2458197.00901 &  23.932888 &  0.374788 &  0.073781 &  \nd \\
 251809768 &  18.310 &  2458182.00880 &  3.744813 &  0.132943 &  0.027276 &  \nd \\
 251809787 &  16.978 &  2458180.14621 &  0.874333 &  0.111146 &  0.174670 &  \nd \\
 251809799 &  18.088 &  2458179.77296 &  0.929420 &  0.101403 &  0.209458 &  \nd \\
 251809801 &  18.209 &  2458180.14037 &  5.424922 &  0.239628 &  0.047817 &  \nd \\
 251809804 &  18.366 &  2458181.02178 &  3.044908 &  0.394803 &  0.336826 &  \nd \\
 251809805 &  18.431 &  2458179.87263 &  0.493215 &  0.072998 &  0.260563 &  contact \\
 251809808 &  18.531 &  2458179.64709 &  0.986293 &  0.204333 &  0.341796 &  \nd \\
 251809809 &  18.694 &  2458179.63921 &  0.543684 &  0.081742 &  0.091127 &  contact \\
 251809830 &  19.404 &  2458180.01339 &  0.746323 &  0.081742 &  0.313398 &  \nd \\
 251809968 &  19.390 &  2458179.54579 &  0.622505 &  0.081742 &  0.185758 &  \nd \\
 251810686 &  10.865 &  2458186.24598 &  13.191424 &  0.151051 &  0.012218 &  quintuple system, \cite{rappaport:2016} \\
 251539584 &  10.763 &  2458179.55118 &  1.088222 &   0.045042 & 0.000625 & SB2, blend with 251539609\\
 251539609 &  11.016 &  2458179.55151 &  1.088213 &   0.044667 & 0.000624 & SB2, blend with 251539584
 \enddata\label{tab:eb}
\end{deluxetable}

\clearpage

\begin{deluxetable}{rrrl}
\tabletypesize{\scriptsize}\tablecaption{Other Periodic Variables}\tablehead{
  &    $Kp$ &    $P$ &             \\
  EPIC & [mag] & [d] & comments \\
}\startdata
 212404864 &  17.754 &  0.583854 &  \nd \\
 212416035 &  18.061 &  0.650274 &  \nd \\
 212424629 &  16.018 &  0.651446 &  \nd \\
 212424861 &  17.877 &  0.651436 &  \nd \\
 212425817 &  16.684 &  0.715986 &  RR Lyrae \\
 212426904 &  15.519 &  1.559636 &  \nd \\
 212429810 &  9.835 &  1.751454 &  \nd \\
 212431975 &  12.460 &  0.560643 &  \nd \\
 212433098 &  14.338 &  0.755435 &  \nd \\
 212433328 &  14.893 &  1.155617 &  \nd \\
 212439709 &  14.352 &  0.609047 &  contact? \\
 212440192 &  16.146 &  0.531711 &  \nd \\
 212441076 &  14.847 &  0.528502 &  \nd \\
 212443701 &  16.789 &  0.683153 &  \nd \\
 212449290 &  16.309 &  0.847446 &  \nd \\
 212449840 &  14.091 &  0.558064 &  \nd \\
 212450261 &  12.888 &  3.746695 &  \nd \\
 212453596 &  16.109 &  0.595544 &  \nd \\
 212460039 &  9.020 &  0.571204 &  \nd \\
 212461484 &  7.976 &  2.268343 &  \nd \\
 212463213 &  14.966 &  0.644204 &  \nd \\
 212467265 &  16.591 &  0.617039 &  \nd \\
 212469922 &  12.509 &  0.810722 &  \nd \\
 212470542 &  14.767 &  0.501587 &  \nd \\
 212470959 &  16.904 &  0.909599 &  \nd \\
 212475454 &  14.591 &  0.495057 &  \nd \\
 212476230 &  14.065 &  0.909933 &  \nd \\
 212476743 &  16.906 &  0.626211 &  \nd \\
 212476895 &  12.756 &  0.806344 &  \nd \\
 212478962 &  15.411 &  0.609325 &  \nd \\
 212479061 &  18.334 &  0.491113 &  \nd \\
 212481276 &  14.791 &  0.560738 &  \nd \\
 212491978 &  14.025 &  0.535797 &  \nd \\
 212492961 &  12.942 &  0.746502 &  \nd \\
 212503342 &  8.324 &  0.501263 &  \nd \\
 212504059 &  11.601 &  0.505806 &  \nd \\
 212506921 &  16.857 &  0.537091 &  \nd \\
 212506981 &  18.107 &  0.560708 &  \nd \\
 212519490 &  12.859 &  0.553239 &  \nd \\
 212520127 &  16.474 &  0.787684 &  \nd \\
 212529254 &  15.890 &  1.224833 &  \nd \\
 212530684 &  17.050 &  0.505286 &  large OOT amplitude \\
 212534342 &  17.713 &  0.617741 &  \nd \\
 212537690 &  16.567 &  0.605773 &  \nd \\
 212540092 &  17.920 &  0.558487 &  \nd \\
 212542474 &  12.033 &  0.526188 &  \nd \\
 212551424 &  13.270 &  0.634884 &  \nd \\
 212555590 &  14.733 &  0.636359 &  \nd \\
 212560096 &  14.764 &  0.599002 &  \nd \\
 212561206 &  15.129 &  0.615971 &  \nd \\
 212562145 &  14.856 &  0.728760 &  \nd \\
 212564937 &  14.129 &  0.506676 &  \nd \\
 212570257 &  12.523 &  0.610247 &  \nd \\
 212575000 &  16.145 &  0.735286 &  \nd \\
 212575799 &  15.277 &  0.616666 &  \nd \\
 212575959 &  12.439 &  0.670392 &  \nd \\
 212578200 &  13.144 &  1.131015 &  \nd \\
 212589990 &  12.178 &  0.504842 &  \nd \\
 212594525 &  15.888 &  0.762575 &  \nd \\
 212597328 &  18.187 &  0.658850 &  RR Lyrae \\
 212601233 &  14.997 &  0.636031 &  \nd \\
 212603282 &  12.328 &  0.696329 &  \nd \\
 212603536 &  11.933 &  0.720349 &  \nd \\
 212603999 &  15.443 &  0.502387 &  RR Lyrae \\
 212609833 &  16.543 &  0.570110 &  \nd \\
 212612729 &  14.534 &  0.904916 &  \nd \\
 212617685 &  13.406 &  0.594009 &  \nd \\
 212619206 &  15.542 &  0.687767 &  \nd \\
 212620826 &  13.616 &  0.789620 &  \nd \\
 212621423 &  14.951 &  0.817041 &  \nd \\
 212628986 &  15.071 &  1.428411 &  \nd \\
 212631286 &  13.236 &  0.525008 &  \nd \\
 212631414 &  13.022 &  0.525005 &  \nd \\
 212631757 &  16.082 &  0.175266 &  \nd \\
 212636050 &  15.543 &  0.630885 &  \nd \\
 212639395 &  16.928 &  0.591004 &  \nd \\
 212639932 &  16.316 &  0.619463 &  \nd \\
 212640806 &  15.889 &  0.510041 &  \nd \\
 212642195 &  14.144 &  0.629391 &  \nd \\
 212644219 &  16.174 &  0.622971 &  \nd \\
 212648945 &  13.771 &  0.750334 &  \nd \\
 212659834 &  11.665 &  0.546711 &  \nd \\
 212666537 &  16.115 &  0.494617 &  \nd \\
 212669531 &  13.967 &  0.606174 &  \nd \\
 212672666 &  16.536 &  0.520714 &  \nd \\
 212674862 &  15.842 &  0.675189 &  \nd \\
 212676658 &  10.640 &  0.532304 &  \nd \\
 212699845 &  17.389 &  0.616183 &  \nd \\
 212703179 &  11.251 &  0.673494 &  \nd \\
 212704410 &  10.588 &  0.762124 &  \nd \\
 212706992 &  14.171 &  0.573939 &  \nd \\
 212711185 &  15.760 &  0.676885 &  \nd \\
 212711671 &  14.949 &  0.545729 &  \nd \\
 212715425 &  14.822 &  0.542155 &  \nd \\
 212716271 &  15.192 &  0.546693 &  \nd \\
 212716448 &  18.478 &  0.546688 &  \nd \\
 212716631 &  18.970 &  0.573803 &  \nd \\
 212717166 &  16.262 &  0.586327 &  \nd \\
 212718800 &  13.631 &  0.650108 &  \nd \\
 212719030 &  15.126 &  1.349336 &  \nd \\
 212720186 &  16.530 &  0.626749 &  \nd \\
 212722087 &  12.587 &  0.546000 &  \nd \\
 212722872 &  14.345 &  0.692869 &  \nd \\
 212723581 &  15.961 &  0.600851 &  \nd \\
 212730754 &  17.858 &  0.587020 &  \nd \\
 212732420 &  13.805 &  0.546859 &  \nd \\
 212733211 &  16.553 &  0.592465 &  \nd \\
 212735753 &  17.112 &  0.611941 &  \nd \\
 212736684 &  18.155 &  0.548902 &  \nd \\
 212742333 &  18.142 &  0.582756 &  \nd \\
 212749368 &  16.551 &  0.630246 &  \nd \\
 212755404 &  13.810 &  0.758773 &  \nd \\
 212760038 &  11.199 &  0.598949 &  \nd \\
 212766036 &  16.427 &  1.128395 &  \nd \\
 212775050 &  16.256 &  0.633570 &  \nd \\
 212775136 &  13.127 &  0.520693 &  \nd \\
 212783579 &  13.453 &  0.623693 &  \nd \\
 212784817 &  15.000 &  0.735008 &  \nd \\
 212785152 &  15.295 &  0.688545 &  \nd \\
 212791551 &  19.214 &  0.720158 &  \nd \\
 212791701 &  16.337 &  0.533695 &  \nd \\
 212793961 &  12.154 &  0.633511 &  \nd \\
 212794694 &  17.778 &  0.505073 &  \nd \\
 212794999 &  16.022 &  0.602511 &  \nd \\
 212795516 &  17.724 &  0.613296 &  \nd \\
 212798939 &  16.823 &  0.507892 &  \nd \\
 212801998 &  15.450 &  0.517430 &  \nd \\
 212808944 &  13.005 &  0.670074 &  \nd \\
 212812050 &  13.882 &  0.575880 &  \nd \\
 212814000 &  14.807 &  0.561011 &  \nd \\
 212814419 &  18.297 &  0.625019 &  \nd \\
 212814441 &  14.201 &  0.783737 &  \nd \\
 212818222 &  16.219 &  0.584496 &  \nd \\
 212818294 &  16.194 &  0.829784 &  \nd \\
 212820594 &  14.665 &  0.530704 &  \nd \\
 212821516 &  11.946 &  0.508947 &  \nd \\
 212824416 &  16.638 &  0.590808 &  \nd \\
 212827294 &  16.930 &  0.559323 &  \nd \\
 212828640 &  14.934 &  0.592274 &  \nd \\
 212828933 &  14.283 &  0.716170 &  \nd \\
 212829102 &  12.264 &  0.500330 &  \nd \\
 212829130 &  16.467 &  0.646563 &  \nd \\
 212829294 &  17.079 &  0.754500 &  \nd \\
 212830414 &  16.810 &  0.571236 &  \nd \\
 212831062 &  15.007 &  0.705463 &  \nd \\
 212831234 &  13.076 &  0.649151 &  \nd \\
 212833004 &  9.158 &  0.543036 &  \nd \\
 212835551 &  12.676 &  0.562135 &  \nd \\
 212835780 &  16.332 &  1.673125 &  \nd \\
 212847938 &  15.743 &  0.607034 &  \nd \\
 212853330 &  16.549 &  0.587536 &  \nd \\
 212862638 &  15.191 &  0.497067 &  \nd \\
 212867164 &  17.189 &  0.572633 &  \nd \\
 212869088 &  17.220 &  0.505407 &  \nd \\
 212870977 &  14.714 &  0.507252 &  \nd \\
 212873395 &  12.808 &  0.605284 &  \nd \\
 212879205 &  12.829 &  0.649341 &  \nd \\
 212879653 &  11.576 &  0.517211 &  \nd \\
 212881555 &  17.099 &  0.545534 &  \nd \\
 212882485 &  15.839 &  0.624794 &  \nd \\
 212882871 &  19.921 &  0.612855 &  \nd \\
 212883764 &  15.503 &  0.668488 &  \nd \\
 212884307 &  13.143 &  0.583500 &  \nd \\
 229228086 &  17.360 &  0.620306 &  \nd \\
 229228087 &  17.630 &  0.602832 &  \nd \\
 229228091 &  18.240 &  0.600837 &  \nd \\
 229228112 &  17.940 &  0.591997 &  \nd \\
 229228121 &  17.770 &  0.574762 &  \nd \\
 251501619 &  14.964 &  0.580914 &  \nd \\
 251502557 &  13.714 &  0.679484 &  \nd \\
 251504831 &  17.611 &  0.622515 &  \nd \\
 251504891 &  9.777 &  0.528140 &  \nd \\
 251505259 &  17.675 &  0.622474 &  \nd \\
 251509348 &  16.172 &  0.623298 &  \nd \\
 251517127 &  18.061 &  0.714932 &  \nd \\
 251519864 &  11.446 &  1.275710 &  \nd \\
 251520093 &  18.417 &  0.540185 &  \nd \\
 251523672 &  16.201 &  0.594779 &  \nd \\
 251526009 &  18.424 &  0.672721 &  \nd \\
 251529654 &  16.234 &  0.521895 &  \nd \\
 251530257 &  17.204 &  0.641235 &  \nd \\
 251540409 &  16.770 &  0.537995 &  \nd \\
 251554210 &  16.357 &  0.509245 &  \nd \\
 251564868 &  18.244 &  0.494339 &  \nd \\
 251566981 &  11.096 &  0.518554 &  \nd \\
 251568443 &  14.911 &  0.714645 &  \nd \\
 251569406 &  14.271 &  0.670480 &  \nd \\
 251574051 &  13.248 &  2.206687 &  \nd \\
 251578582 &  11.275 &  7.120210 &  \nd \\
 251579007 &  14.922 &  0.629344 &  \nd \\
 251583296 &  17.090 &  0.549769 &  \nd \\
 251583388 &  14.011 &  0.950893 &  \nd \\
 251585662 &  19.180 &  0.646642 &  \nd \\
 251590688 &  12.081 &  0.710497 &  \nd \\
 251596880 &  10.890 &  2.633147 &  \nd \\
 251599500 &  15.101 &  0.571171 &  \nd \\
 251602987 &  17.865 &  0.688673 &  \nd \\
 251608983 &  12.951 &  0.934933 &  \nd \\
 251611842 &  12.691 &  0.518191 &  \nd \\
 251612403 &  15.626 &  0.698081 &  \nd \\
 251613106 &  17.050 &  0.717477 &  \nd \\
 251615995 &  14.797 &  0.561389 &  \nd \\
 251809762 &  17.770 &  0.574708 &  \nd \\
 251809767 &  18.290 &  0.609255 &  \nd \\
 251809792 &  17.702 &  0.582034 &  \nd \\
 251809793 &  17.830 &  0.535073 &  \nd \\
 251809794 &  17.837 &  0.514385 &  \nd \\
 251809800 &  18.158 &  0.644357 &  \nd \\
 251809802 &  18.232 &  0.565049 &  \nd \\
 251809803 &  18.271 &  0.538007 &  \nd \\
 251809807 &  18.499 &  0.605395 &  \nd \\
 251809812 &  18.954 &  0.615473 &  \nd \\
 251809817 &  19.009 &  0.598227 &  \nd \\
 251809820 &  19.110 &  0.573687 &  \nd \\
 251809824 &  19.182 &  0.709409 &  \nd \\
 251809836 &  19.611 &  0.591795 &  \nd \\
 251809865 &  20.310 &  0.669433 &  \nd \\
 251810875 &  18.667 &  0.643312 &  \nd \\
 251811189 &  18.981 &  0.560705 &  \nd \\
 251811486 &  19.100 &  0.798840 &  \nd \\
 251811829 &  19.187 &  0.651565 &  \nd \\
 251809821 &  19.110 &  0.610251 &  \nd \\
\enddata\label{tab:var}
\end{deluxetable}

\end{document}